%% file: manuscript.tex
\documentclass[manuscript]{acmart}
\AtBeginDocument{%
  }

\input{shortcuts}
\usepackage{multirow}
\usepackage{adjustbox}
\begin{document}
\begin{teaserfigure}
  \centering
  \includegraphics[width=\linewidth]{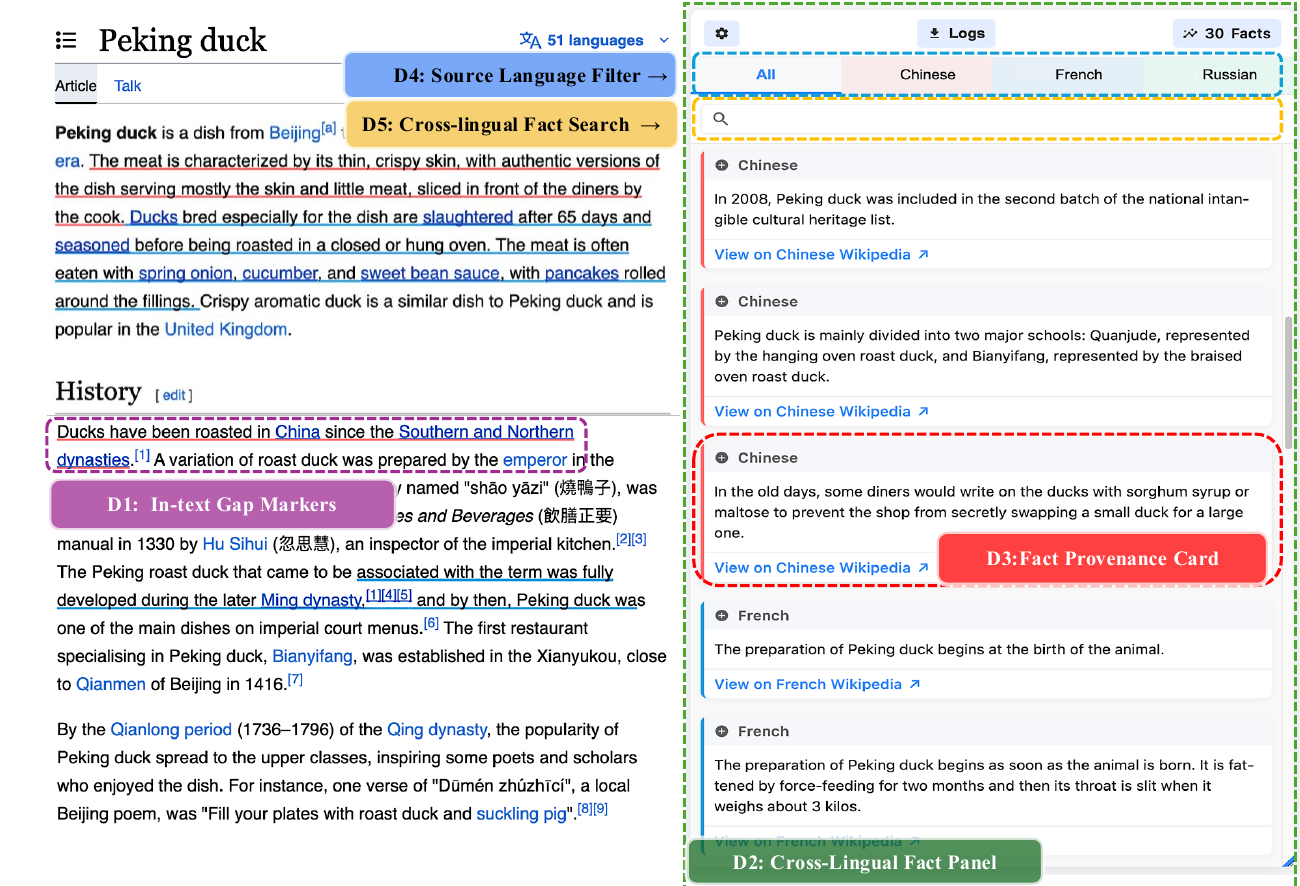}
  \caption{The \textsc{WikiGap} interface embeds cross-lingual facts into English Wikipedia via five key design elements (D1–D5), supporting in-place access, traceability, and multilingual engagement.}
  \label{fig:ui}
\end{teaserfigure}
% \title{\textsc{WikiGap}: A Chrome Extension for Surfacing 
% Cross-Lingual Knowledge Gaps in Wikipedia}

\title[Surfacing Knowledge Gaps Between English Wikipedia and other Language Editions]{WikiGap: Promoting Epistemic Equity by Surfacing Knowledge Gaps Between English Wikipedia and other Language Editions} 


% \affiliation{%
\author{Zining Wang}
\affiliation{%
  \institution{University of British Columbia}
  \country{Canada}
}
\affiliation{%
  \institution{Vector Institute for AI}
  \country{Canada}
}
\email{zining.wang@ubc.ca}

\author{Yuxuan Zhang}
\affiliation{%
  \institution{University of British Columbia}
  \country{Canada}
}

\author{Dongwook Yoon}
\affiliation{%
  \institution{University of British Columbia}
  \country{Canada}
}

\author{Nicholas Vincent}
\affiliation{%
  \institution{Simon Fraser University}
  \country{Canada}
}

\author{Farhan Samir}
\affiliation{%
  \institution{University of British Columbia}
  \country{Canada}
}
\affiliation{%
  \institution{University of Toronto}
  \country{Canada}
}

\author{Vered Shwartz}
\affiliation{%
  \institution{University of British Columbia}
  \country{Canada}
}
\affiliation{%
  \institution{Vector Institute for AI}
  \country{Canada}
}

\renewcommand{\shortauthors}{Wang et al.}

\input{sections/0-abstract}
\maketitle

\section{Introduction} \label{section:introduction}
\input{revised-sections/1-introduction-revised}

\section{Background and Related Work} \label{section:related-work}
\input{revised-sections/2-related-work-revised}

\section{WikiGap System} \label{section:implementation}
\input{revised-sections/3-implementation-revised}

\section{Evaluation Methods} \label{sec:eval:method}
\input{revised-sections/4-evaluation-revised}

\section{Findings}
\label{sec:findings}
\input{revised-sections/5-findings-revised}

\section{Discussion} \label{section:discussion}
\input{revised-sections/6-discussion-revised}

\section{Conclusion} \label{section:conclusion}
\input{sections/7-conclusion}

% \section{Acknowledgments}
% ChatGPT was utilized to generate code snippets and to assist in refining the grammar and writing style of this work. All AI-generated content was reviewed and edited by the authors to ensure accuracy and adherence to academic standards.

\begin{acks}
    ChatGPT was utilized to generate code snippets and to assist in refining the grammar and writing style of this work. All AI-generated content was reviewed and edited by the authors to ensure accuracy and adherence to academic standards.
    This work was funded, in part, by the Vector Institute for AI, Canada CIFAR AI Chairs program, Accelerate Foundation Models Research Program Award from Microsoft, and an NSERC discovery grant
\end{acks}

\bibliographystyle{ACM-Reference-Format}
\bibliography{references}

\input{sections/appendix}

\end{document}

%% file: shortcuts.tex
\newcommand{\tocheck}[1]{\textcolor{purple}{#1}}

\newif\ifmarkchanges
\markchangesfalse         % <-- CLEAN version (uncomment for clean)

\ifmarkchanges
  \DeclareRobustCommand{\tocheck}[1]{\textcolor{purple}{#1}}
\else
  \DeclareRobustCommand{\tocheck}[1]{#1}
\fi

\newif\ifshowchanges
\ifshowchanges
   \usepackage{xcolor}
   \usepackage[normalem]{ulem}
   \newcommand{\del}[1]{\textcolor{red}{\sout{#1}}}

\else
   \newcommand{\del}[1]{}

\fi

%% file: sections/0-abstract.tex
\begin{abstract}
With more than 11 times as many pageviews as the next largest edition, English Wikipedia dominates global knowledge access relative to other language editions. Readers are prone to assuming English Wikipedia as a superset of all language editions, leading many to prefer it even when their primary language is not English. Other language editions, however, comprise complementary facts rooted in their respective cultures and media environments, which are marginalized in English Wikipedia. While Wikipedia's user interface enables switching between language editions through its Interlanguage Link (ILL) system, it does not reveal to readers that other language editions contain valuable, complementary information. We present \textsc{WikiGap}, a system that surfaces complementary facts sourced from other Wikipedias within the English Wikipedia interface. Specifically, by combining a recent multilingual information-gap discovery method \citep{samir-etal-2024-locating} with a user-centered design, \textsc{WikiGap} enables access to complementary information from French, Russian, and Chinese Wikipedia. In a mixed-methods study (n=21), \textsc{WikiGap} significantly improved fact-finding accuracy, reduced task time, and received a 32-point higher usability score relative to Wikipedia's current ILL-based navigation system. Participants reported increased awareness of the availability of complementary information in non-English editions and reconsidered the completeness of English Wikipedia. \textsc{WikiGap} thus paves the way for promoting epistemic equity across language editions. Taken together, these findings show how interface design can redistribute reader attention and engagement across Wikipedia’s multilingual knowledge production communities, supporting epistemic equity.
\end{abstract}

%% file: revised-sections/1-introduction-revised.tex
Wikipedia hosts more than 300 language editions, inviting contributions from many distinct linguistic and geographic regions. The investment in creating multiple language editions reflects Wikipedia's principles of democratizing both access to and contribution of information. These language editions are sociogeographically situated; it is generally well understood in Wikipedia scholarship that they are not mere translations of one another \citep{hecht2013mining}. Instead, the language editions tend towards reflecting the knowledge that is contextually relevant, often based on cultural experience and localized media diets \citep{avieson2025wiki,kumar2021digital}.

However, this heterogeneity is not well understood by the broader public that engage with Wikipedia, whether readers or contributors \citep{miquel2018wikipedia}. Rather, the popular understanding of multilingual Wikipedia is that there is a global consensus on what information is reliable and noteworthy \citep{hecht2013mining}. Moreover, English Wikipedia is by and large perceived as the most comprehensive and complete account of this consensus, containing as much as or more information than any other, making other languages the minority. This \textit{English-as-superset} assumption is widespread, even among people who speak multiple languages proficiently; in a 2011 report, $93\%$ of all editors surveyed attested to reading English Wikipedia, even though it was a primary language for only $52\%$ of them \citep{kumar2021digital,wikipedia2011editorsurvey}. In the same vein, $20\%$ of English Wikipedia's visitors were based in India, even more than readers from the United Kingdom. English Wikipedia is so widely understood to be an epistemic authority that is commonly treated as a source of neutral and complete information in and of itself by editors contributing to minority language editions \citep{hickman2021understanding}.

This epistemic assumption is consequential not only for readers' understanding of Wikipedia but also for the organization of its peer production communities. Wikipedia is sustained by interconnected yet unevenly resourced volunteer communities, each centered on different language editions \citep{sen2015barriers}. Prior research has shown that volunteer engagement in Wikipedia depends on motivational factors such as enjoyment, opportunities for learning, and perceived recognition within a contributor community, beyond shared ideological commitments to free and open knowledge \citep{Nov2007, Arazy2017}. When English Wikipedia is widely perceived as the most complete and authoritative source, reader attention and contributor effort become disproportionately concentrated in the English-language community. This concentration structurally amplifies the conditions that prior work has shown to support sustained participation, while making those conditions less accessible in minority language editions. One byproduct of this imbalance is a reduction in incentives for volunteer contributions to those communities \citep{Maiberg2025}, undermining Wikipedia’s ethos as a globally and cooperatively produced multilingual knowledge infrastructure. In this work, we ask how interface design can redistribute attention and engagement across Wikipedia’s multilingual peer production communities, promoting epistemic equity. \footnote{Throughout this paper, we use epistemic equity as a design lens, focusing specifically on how interface design can redistribute attention and engagement across Wikipedia’s multilingual peer production communities, rather than addressing broader institutional or political dimensions of epistemic equity.}

% This epistemic assumption is consequential, as its perception as an objective and complete knowledge base monopolizes attention away from the prospect of unique perspectives in other language editions. Specifically the peer production community for English Wikipedia monopolizes a lot of attention and contributions. One byproduct of this monopolization is that it reduces incentives for volunteer contributions to those minority language editions \citep{Maiberg2025}. The \textit{English-as-superset} assumption then runs counter to Wikipedia's ethos of a global democratically-curated multilingual knowledge base. In this work, we ask how can we counter this epistemic assumption through a design, thereby promoting epistemic equity? %How can design challenge English Wikipedia’s assumed universality and support epistemic equity across languages?

Prior works have sought to incorporate contributions in minority language editions by consolidating information from all editions into a single unified knowledge base \citep[e.g.,][]{duh2013managing,adar2009information}. However, this conceptual formulation fails to \textit{center} minority language editions and the unique volunteer contributions they contain. It continues to perpetuate the ideology of there being a global consensus on noteworthy and reliable information, a universal knowledge base. As such, it does little to challenge the status quo of English Wikipedia as an objective and complete epistemic authority. 

In our work we take a social constructionist view of knowledge \citep{haraway2013situated} in approaching the problem by improving engagement with minority language editions. That is, we center necessarily incomplete and sociogeographically positioned curation of knowledge. To highlight the positionality (rather than objectivity) of English Wikipedia, we aim to highlight its content gaps relative to other language editions. At the same time, we balance this intervention against being minimally disruptive to the current Wikipedia reading design \citep{Grudin1988}. We present \textsc{WikiGap}, an augmentation of the Wikipedia reading experience that surfaces sentence-level factual differences from English to 3 other language editions of Wikipedia (French, Russian, and Chinese). The system combines a computational backend (\textsc{InfoGap}) that identifies multilingual content gaps~\cite{samir-etal-2024-locating} with a browser-based interface that presents these differences through in-page highlights and a sidebar display (Figure~\ref{fig:ui}). 

We evaluated \textsc{WikiGap} through a mixed-methods user study using a fact-finding task focused on culturally rich food articles -- a domain where cross-lingual disparities are particularly salient~\citep{laufer2015mining}. Our results show that participants strongly preferred \textsc{WikiGap} over Wikipedia's default interlanguage links system (ILLs), with System Usability Scale (SUS) scores 32 points higher on average. In terms of performance, participants were significantly more accurate and faster at answering factual questions when using \textsc{WikiGap}, suggesting that surfacing multilingual information directly within the English interface not only improves user satisfaction but also enhances learning efficiency.

Through a Post-Study interview, we find that participants, prior to using \textsc{WikiGap}, were largely unaware of content gaps in English Wikipedia, demonstrating that the problematic \textit{English-as-superset} epistemic assumption persists more than a decade after it was identified in seminal studies by \citet{hecht2013mining}. 
% highlighting the importance of raising awareness of the multilingual composition of Wikipedia to the broader public. 
% Users reported a shift in their perception of Wikipedia's completeness, expressing surprise at how much content in other language editions was missing from English Wikipedia. 
Participants were overwhelmingly appreciative of \textsc{WikiGap}'s capacity to surface these substantive information gaps during the reading of English Wikipedia pages, and expressed openness and enthusiasm about interacting with multilingual content in English Wikipedia.  
Moreover, a few participants with prior editing experience described how the system could support their editorial workflows by enabling source verification and facilitating the integration of multilingual content. These findings underscore \textsc{WikiGap}'s broader value as a bridge between communities of editors and readers across languages. 

In Section~\ref{section:discussion}, we reflect on how \textsc{WikiGap} challenges the English-as-superset assumption by redistributing reader attention and engagement across Wikipedia’s multilingual peer production communities. We further interpret the system as a boundary object \citep{star1989institutional,susan2010} that supports articulation work between readers and editors across language editions. We contribute a validated system design for surfacing fact-level cross-lingual information gaps in Wikipedia. The system combines a computational pipeline that detects and integrates missing facts from non-English editions with an unobtrusive interface for English readers. A mixed-methods evaluation (n=21) shows that this design improves fact-finding accuracy and efficiency, outperforms Wikipedia’s interlanguage-link system in usability, and increases engagement with multilingual knowledge—challenging assumptions about English Wikipedia’s epistemic completeness.\footnote{Code and data will be made publicly available upon publication.}  

% This 
% paper makes two key contributions. First, we present a novel system design that leverages a computational pipeline to detect fact-level information gaps across languages and integrate relevant facts from other language editions directly into English Wikipedia articles. Second, through an empirical mixed-methods study (N=21), we demonstrate that this system unobtrusively introduces multilingual knowledge to readers, significantly improving fact-finding accuracy and completion time, achieving higher usability than Wikipedia’s default interlanguage-link system, and fostering greater engagement with multilingual content—thereby challenging the perceived epistemic supremacy of English Wikipedia.

%% file: revised-sections/2-related-work-revised.tex
Wikipedia's multilingual architecture aspires to democratize knowledge access across languages, yet its design and community practices systematically privilege English content. In \S\ref{sec:bg:epistemic}, we examine how this ``English-as-superset" assumption \citep{hecht2013mining} persists despite empirical evidence to the contrary, and how current Wikipedia interface reinforces epistemic hierarchies that undermine the platform's multilingual vision by concentrating attention and participation within a single language edition. In \S\ref{sec:bg:prior}, we explain that cross-linguistic variation should be surfaced rather than obscured by drawing literature from prior CSCW research on asset-based design \citep{10.1145/3406865.3418594, kretzmann1993building, 10.1145/3637291} and feminist epistemology \citep{haraway2013situated, bardzell2010feminist}. Then, we review prior technical interventions designed to surface cross-linguistic knowledge differences, evaluating them along two critical dimensions: the granularity of differences they expose and their integration with readers' existing browsing practices to support engagement across language communities. This analysis reveals persistent limitations that motivate our \textsc{WikiGap} design.

\subsection{English-as-Superset and Epistemic Inequity in Multilingual Wikipedia}
\label{sec:bg:epistemic}

Wikipedia embodies a central tension in collaborative knowledge systems: although it is architecturally multilingual and aspires to provide knowledge to ``every single person on the planet in their own language'' \citep{Cohen2008}, its design and community practices systematically privilege English content \citep{foucault2013archaeology,said1977orientalism}. This imbalance shapes how knowledge is organized and encountered on the platform, influencing both which perspectives become visible and how reader attention is distributed across language-specific peer production communities. At the interface level, this dynamic obscures cross-linguistic differences and reinforces what \citet{hecht2013mining} term the ``English-as-superset'' assumption—the belief that English Wikipedia exhaustively represents the knowledge available across other language editions.

CSCW scholarship on epistemic equity \citep{ajmani2024whose, benjamin2023race} helps explain why this assumption matters: collaborative systems should support the visibility and engagement of diverse epistemic standpoints rather than subsuming them under a dominant perspective. In Wikipedia’s multilingual context, this means recognizing that language communities produce knowledge shaped by distinct cultural, geographic, and historical conditions that are not uniformly visible across language editions

\subsubsection{English Wikipedia Is Treated as the Default, Yet Knowledge Gaps Persist} Given English’s status as a global lingua franca, the English edition of Wikipedia has far outpaced other language editions in size and coverage. As a result, it is often implicitly treated as a superset of their knowledge. \citet{hecht2013mining} identified this ``English-as-superset'' viewpoint, which, though rarely stated explicitly, remains prevalent—even among researchers who rely on multilingual Wikipedia to train text-generation and text-embedding models. However, substantial knowledge gaps persist between English and other language editions. \citet{hecht2010tower} showed that English Wikipedia lacks coverage on many topics documented in other languages, while more recent work demonstrates that even when topics overlap, English articles often omit facts present elsewhere \citep{samir-etal-2024-locating}.

\subsubsection{From Knowledge Gaps to Epistemic Invisibility} From an epistemic equity perspective, such variation is not an error or deficiency, but an expected outcome of situated knowledge production. One interpretation of Wikipedia’s multilingual vision assumes a language-agnostic set of facts that can be uniformly translated across languages—a view that treats language as a neutral vessel for meaning, abstracted from socio-geographic context \citep[][Chapter~4]{golumbia2009cultural}. Yet much of Wikipedia’s content concerns culturally and geographically situated histories, practices, and identities \citep{wmf2024-most-popular}. Accordingly, language editions are shaped by the communities that produce them \citep{johnson2022considerations} and exhibit a well-documented ``self-focus bias'' toward regionally relevant topics \citep{hecht2009measuring,samir-etal-2024-locating,callahan2011cultural}. This emphasis on positionality echoes feminist epistemological arguments that knowledge is always produced from particular social, cultural, and linguistic standpoints \citep{haraway2013situated}. When such epistemic differences are rendered invisible, they not only obscure situated knowledge but also limit opportunities for readers and editors to encounter and engage with the communities that produce it.

The ``English-as-superset'' assumption can thus contribute to cycles of epistemic invisibility \citep{ajmani2024whose}, in which visibility and legitimacy are recursively reinforced within collaborative systems. On Wikipedia, these dynamics manifest in both editorial and readership practices. Bilingual editors often consult English Wikipedia to determine what content belongs in other language editions, treating it as a reference point \citep{hickman2021understanding}. Wikipedia’s emphasis on verifiability further privileges English-language sources, creating structural barriers for knowledge rooted in other linguistic or cultural contexts \citep{sen2015barriers}. Although multilingual editors can bridge across editions, they constitute only a small fraction of Wikipedia’s editor base \citep{hale2014multilinguals}.

Readers similarly assume that different language editions contain equivalent information (Section~\ref{sec:findings}). As engagement with smaller editions declines, editors in those communities receive less visibility and recognition, weakening key motivations for participation \citep{Nov2007,Arazy2017}. This feedback loop reduces incentives for contribution and undermines Wikipedia’s vision as a collaboratively produced multilingual knowledge infrastructure \citep{Maiberg2025}.

\subsubsection{Interface Design and the Obscuring of Epistemic Difference}
These epistemic hierarchies are not only reflected in community norms and editorial practices, but are also enacted through interface design. Wikipedia primarily supports cross-lingual navigation through Interlanguage Links (ILLs), presented as a drop-down menu on each article. While ILLs enable access to other language editions, the interface provides no cues indicating that articles may differ substantially in content. By framing language switching as a matter of translation rather than epistemic difference, the design implicitly suggests a shared underlying knowledge base, limiting opportunities for readers to engage with the work of other language communities. Yet prior work shows that many unique facts remain distributed across non-local language editions, even for regionally associated topics \citep{samir-etal-2024-locating}. As a result, Wikipedia’s current interface renders epistemic differences structurally invisible to readers, making it difficult to engage with other language editions.

Together, this body of work motivates a focus on design interventions that challenge the ``English-as-superset'' assumption by making cross-linguistic differences visible and supporting engagement across Wikipedia’s multilingual peer production communities.

\subsection{Prior Work Targeting Cross-Lingual Knowledge Dissemination in Wikipedia}
\label{sec:bg:prior}

\subsubsection{Surfacing vs. Consolidating Cross-Linguistic Differences}
Prior work on cross-linguistic variation in Wikipedia has generally followed two contrasting orientations. One line of work treats differences across language editions primarily as obstacles to information access, motivating efforts to consolidate multilingual knowledge into a single, unified representation \citep{duh2013managing, adar2009information}. While such approaches aim to improve coverage and navigability, they implicitly assume that cross-linguistic variation should be minimized, and risk obscuring the distinct perspectives embedded in distributed knowledge production by enforcing uniform representations \citep{schmidt_taking_1992}.

In contrast, a second orientation emphasizes surfacing cross-linguistic differences as meaningful expressions of positionality \citep{haraway2013situated}. From this perspective, variation across language editions is not a shortcoming to be resolved, but an expected and valuable outcome of situated knowledge production. This framing aligns closely with our earlier discussion of epistemic equity, which calls for collaborative systems to recognize and sustain diverse ways of knowing rather than subsume them under a dominant perspective.

When translated into design terms, this orientation corresponds to an asset-based approach \citep{10.1145/3406865.3418594, kretzmann1993building, 10.1145/3637291}. Rather than treating heterogeneity as a problem to be corrected, asset-based design views existing differences as resources for sense-making and learning. Applied to multilingual Wikipedia, this suggests that interfaces should help readers perceive and engage with cross-linguistic differences, instead of rendering them invisible through consolidation or assumed equivalence. Building on this perspective, our work extends prior efforts by exploring how interface design can surface positionality within everyday Wikipedia browsing practices.

\subsubsection{Prior Multilingual Interface Designs on Surfacing Cross-lingual Knowledge Differences.} 

Several systems have attempted to move beyond Wikipedia's default ILLs by providing interfaces that surface content from multiple language editions. Manypedia \citep{massa2012manypedia} displays two language editions side-by-side using machine translation, but it does not explicitly highlight factual inconsistencies or information gaps between them. WikiCompare \citep{roy2022information} takes a step further by identifying topic-level differences. When a user reads an article in one language (e.g., Hindi), the interface highlights subtopics that are discussed in another language edition (e.g., English) but missing in the current version; clicking a highlight redirects the user to the corresponding English section. While informative, WikiCompare still operates at the topic level, rather than identifying specific factual discrepancies across languages. Moreover, neither Manypedia nor WikiCompare formally evaluates how such cross-lingual surfacing affects readers' comprehension, behavior, or information-gathering efficacy.

% The most pathbreaking of these interfaces is Omnipedia \citep{bao2012omnipedia}, which enables readers to get an overarching sense of how different language editions' editor communities conceptualize a concept. While they similarly rely on anchor links as WikiCompare does, their design enables visualizing several different language editions simultaneously and succinctly. This innovative redesign, however, would require users to completely change how they interact with Wikipedia, The Omnipedia workflow requires that readers start by entering a concept in their search bar. This starting point, however, contrasts with how most readers engage with Wikipedia, through information brokering search engines \citep{vincent2021deeper}. As such, this innovative redesign is unlikely to meet widespread adoption. 

The most ambitious multilingual interface to date is Omnipedia \citep{bao2012omnipedia}, which visualizes how different language communities conceptualize a topic by aggregating anchor-link structures across editions. This design allows simultaneous comparison of multiple languages but comes at the cost of requiring users to adopt an entirely new interaction paradigm. Users must access Omnipedia through its dedicated search interface, which contrasts with how most people encounter Wikipedia—via search engines and in-page browsing \citep{vincent2021deeper}. As a result, Omnipedia's workflow is difficult to integrate unobtrusively into readers' existing habits on Wikipedia.

Complementing such reader-centered systems, the \textsc{Wikipedia Diversity Observatory} (WDO) \citep{miquelribe2020diversityobservatory} offers editor-facing dashboards that quantify cultural and geographic content gaps across language editions and suggest missing but culturally relevant articles. However, WDO also operates outside Wikipedia’s native reading and editing interfaces and focuses only on article-level differences—an even coarser granularity than tools such as WikiCompare.

In summary, prior systems demonstrate the value of surfacing cross-lingual heterogeneity, yet they share several infrastructural limitations that restrict their practical impact: (1) they primarily expose topic-level rather than sentence-level differences; (2) they rely on workflows external to the natural reading experience, limiting their ability to integrate multilingual knowledge unobtrusively, As Grudin's classic analysis of groupware failures highlights \citep{Grudin1988}, systems that impose new routines or additional work misaligned with existing practices frequently struggle to achieve sustained adoption; and (3) with the exception of Omnipedia, they lack empirical evaluation of how such designs actually affect readers.

\subsubsection{Advances in multilingual factual knowledge retrieval}
One of the challenges faced by prior systems that work at the article or topic-level was lack of access to computational tools that can recognize finer-grained differences between articles. Recent advances in multilingual NLP enable automatic detection of sentence-level factual gaps, which we leverage for developing our user-centric \textsc{WikiGap} design. Specifically, we rely on \textsc{InfoGap} \cite[][Sec.~\ref{subsec:infogap}]{samir-etal-2024-locating}. The \textsc{InfoGap} paper corroborated through a large-scale analysis that English Wikipedia lacks facts from other language editions, including facts that may have a broader appeal. For example, Apple CEO Tim Cook's association with the Russia-Ukraine war is only mentioned on Russian Wikipedia, but is conceivably noteworthy to the wider array of readers that engage with English Wikipedia. Building on \textsc{InfoGap}, our \textsc{WikiGap} system embeds those unique facts within Wikipedia's native interface, overcoming limitations from the aforementioned interfaces.

%% file: revised-sections/3-implementation-revised.tex
The goal of \textsc{WikiGap} is to challenge the English-as-superset assumption by making complementary knowledge from other language editions visible to readers during everyday Wikipedia use. Rather than treating multilingual differences as gaps to be corrected, \textsc{WikiGap} is designed to surface them in context, allowing readers to encounter cross-lingual knowledge while browsing English Wikipedia pages. We combine a computational pipeline that identifies sentence-level knowledge gaps across languages with a reader-facing interface that presents these differences through in-situ annotations and structured fact cards.

We begin with outlining the design requirements informed by user needs and foundational theories that shaped the WikiGap interface (\S\ref{sec:design-ui}). We then step behind the scenes to introduce \textsc{InfoGap} \citep{samir-etal-2024-locating}, which we use to detect language-exclusive facts lacking in English Wikipedia (\S\ref{subsec:infogap}). Furthermore, we describe how we extended and adapted \textsc{InfoGap} to detect and deliver language-exclusive facts, including how we filter, translate, and bind them to UI components at runtime (\S\ref{subsec:adapt_infogap}). We conclude this section by outlining the system implementation details (\S\ref{subsec:implementation}).

\input{revised-sections/3-subsec-design-revised}

\subsection{The \textsc{InfoGap} Pipeline}
\label{subsec:infogap}

While the interface makes multilingual differences visible to readers, \textsc{InfoGap} provides the backend capability that enables such differences to be identified at sentence-level granularity. It was a previously developed computational pipeline introduced by Samir et al.~\cite{samir-etal-2024-locating}, which detects factual misalignments between Wikipedia articles across languages.  This subsection introduces the core components of the original \textsc{InfoGap} system, which we use to identify sentence-level knowledge gaps between English and other language editions. In the following section, we will describe how we extended this pipeline to support additional languages and adapted its outputs for integration into the \textsc{WikiGap} interface.

\begin{figure}[t]
  \centering
  \includegraphics[width=0.5\linewidth]{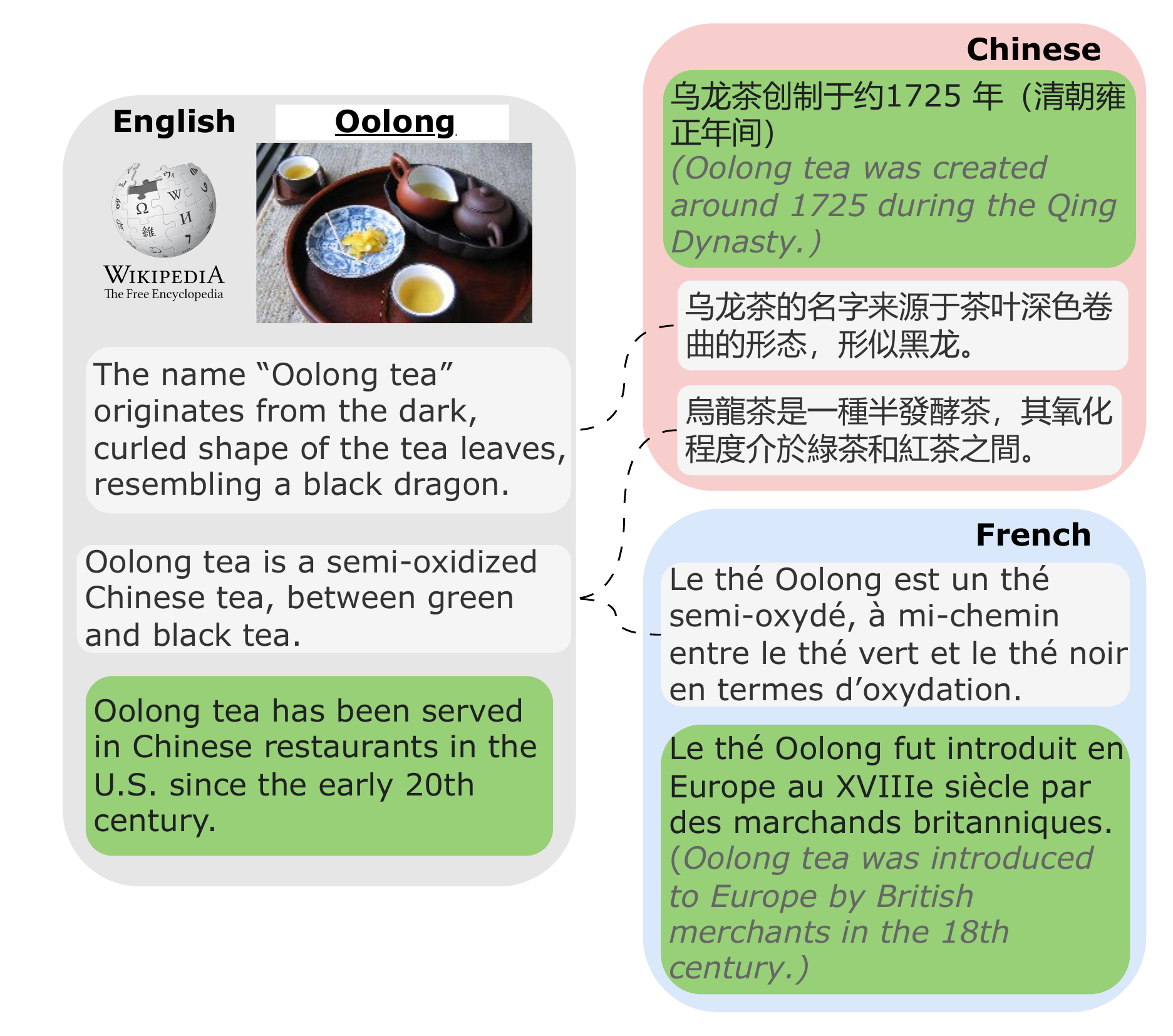}
\caption{Knowledge differences in the Wikipedia coverage of \emph{Oolong} identified by \textsc{InfoGap}. Connecting lines mark overlapping facts; green boxes highlight facts unique to each language edition. English sentences in \textit{italics} represent translations of facts from Chinese or French Wikipedia, while non-italicized English sentences are from the original English Wikipedia article.}
  \label{fig:oolong}
\end{figure}

\textsc{InfoGap} is a state-of-the-art LLM-based pipeline that takes an article $L_s$ in the \emph{source language} and the respective article $L_t$ in the \emph{target language} and returns the sets of common facts and facts that are exclusive to one article. As illustrated in Figure~\ref{fig:oolong}, given a Wikipedia article in the source language ($L_s$, e.g., English) and the respective article in another language ($L_t$, e.g., French), 
% English serves as the source language ($L_s$), and either French or Chinese serves as the target language ($L_t$). 
\textsc{InfoGap} identifies three categories of facts: those shared across editions, those unique to the source, and those unique to the target (i.e., language-exclusive facts). For example, ``\textit{Oolong is a semi-oxidized Chinese tea}'' is shared across languages, while ``\textit{It is served in US restaurants}'' may appear only in English.

\begin{figure}
  \centering
  \includegraphics[width=\linewidth]{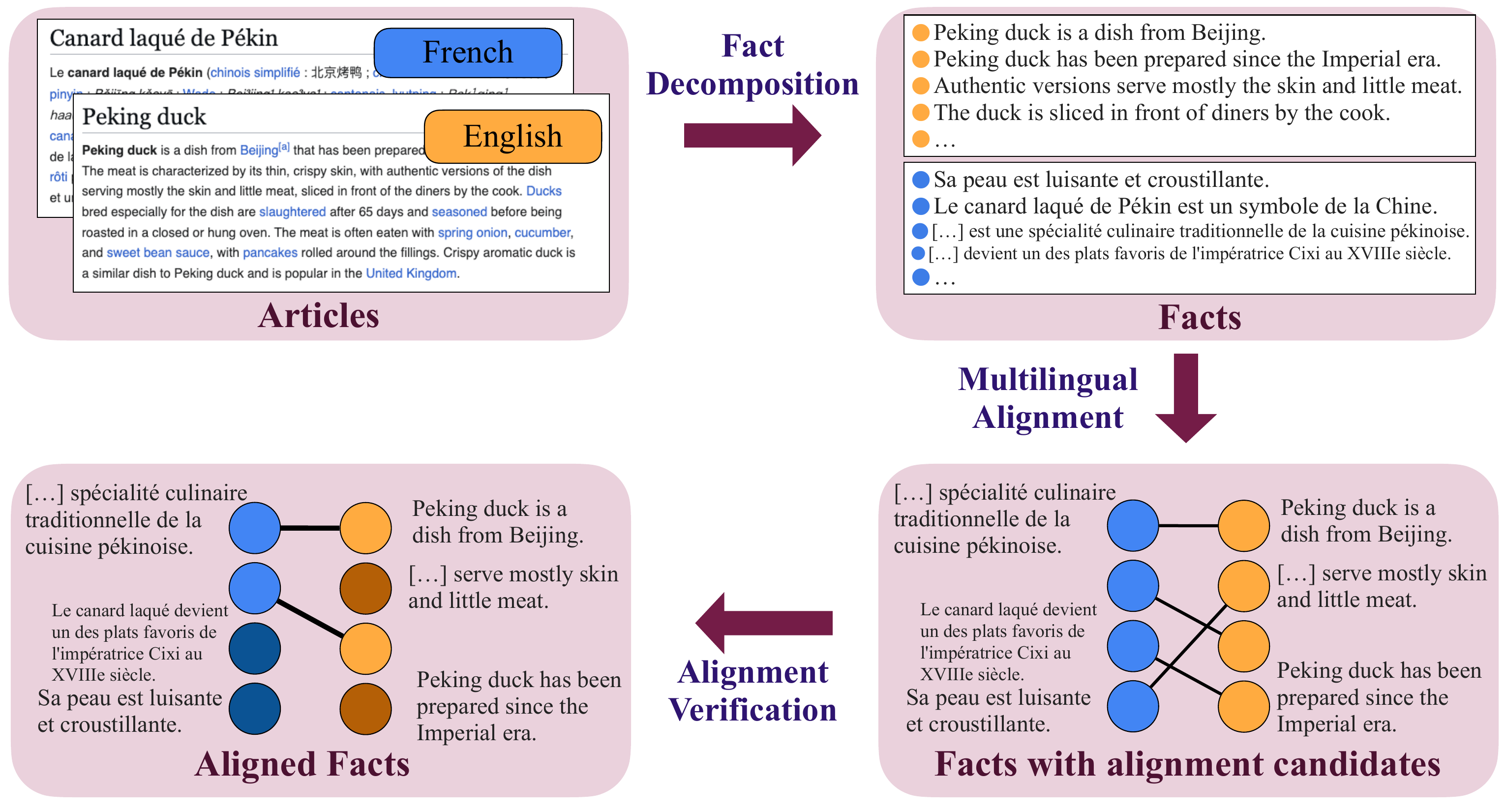}
  \caption{Overview of the \textsc{InfoGap} backend pipeline for cross-lingual fact alignment, reproduced from \citet{samir-etal-2024-locating} and adapted. For additional technical details, see the original paper.}
  \label{fig:infogap}
\end{figure}

The pipeline leverages a multilingual LLM and sentence embeddings to perform cross-lingual comparison in three stages, as illustrated in Figure~\ref{fig:infogap}:

\begin{enumerate}
    \item \textbf{Fact decomposition}.  
    Each paragraph in $L_s$ and $L_t$ is decomposed into atomic factual statements using prompts issued to a multilingual LLM. The output includes both the fact and its paragraph index for downstream alignment.

    \item \textbf{Multilingual alignment}.  
    Each fact is encoded using a multilingual sentence encoder. For every fact in $L_t$, the top three nearest neighbors in $L_s$ are retrieved based on cosine similarity. This reduces noise and narrows the search space when looking for equivalent facts. % scope to likely matches.

    \item \textbf{Alignment verification}.  
    For each fact in $L_t$, the LLM is prompted to determine whether it is inferable from any of its nearest neighbors in $L_s$. If a matching fact is found in $L_s$, these facts are considered aligned, otherwise, the fact is labeled as a \emph{knowledge gap} in $L_s$.
\end{enumerate}

% (2) the integration layer, which maps \textsc{InfoGap} outputs into \textsc{WikiGap}'s data files and interface.

\subsection{Extending and Integrating \textsc{InfoGap} for \textsc{WikiGap}}
\label{subsec:adapt_infogap}
In this section, we outline the conceptual modifications and integration steps that enable \textsc{InfoGap} to serve as the backend for \textsc{WikiGap}. %For technical and implementation-specific details, see Section~\ref{subsec:implementation}. 
We adapted and extended the existing \textsc{InfoGap} pipeline \cite{samir-etal-2024-locating} in two key ways: (1) by enabling new language (Chinese) support beyond the original study, and (2) by building an integration layer that transforms \textsc{InfoGap}'s output into interactive interface components usable within our system.

% \subsubsection{Enabling Chinese language support.}
% The original \textsc{InfoGap} pipeline was developed for detecting cross-lingual gaps in biographical articles between English and either Russian or French. To generalize the method to a new domain and an additional language, we incorporated support for Chinese into the fact decomposition and verification modules. This involved custom preprocessing for tokenization, crafting prompts, and conducting human evaluation on Chinese texts to confirm that \textsc{InfoGap} performs comparably to its performance on the original languages studied by \citet{samir-etal-2024-locating}. %ensuring sentence-level semantic embeddings remained accurate across all three target languages. 
% % These modifications allow \textsc{WikiGap} to surface knowledge from Chinese Wikipedia in addition to French and Russian.

\subsubsection{Rationale for Language Selection}
In this study, we intentionally focused on English and three additional language editions, namely  Russian, French, and Chinese. We chose the first two because the original \textsc{InfoGap} pipeline was developed for detecting cross-lingual gaps in biographical articles between English and either Russian or French. We included them because of they have been validated in the research to ensure accuracy. To generalize the method to a new domain and an additional language, we incorporated support for Chinese into the fact decomposition and verification modules, as some co-authors were proficient in Chinese. This involved custom preprocessing for tokenization, crafting prompts, and conducting human evaluation on Chinese texts to confirm that \textsc{InfoGap} performs comparably to its performance on the original languages studied by \citet{samir-etal-2024-locating}.

\subsubsection{Language attribution and fact selection}
\label{sec:wikigap:integration:language_fact}
Our focus is on \textit{knowledge gaps} -- facts that appear in the $L_t$ version (e.g., French, Russian, or Chinese) of an article but \emph{not} in $L_s$ (English). We therefore filter the \textsc{InfoGap} output to retain only gap facts. To ensure interface responsiveness, we use a precomputed output from \textsc{InfoGap} for each article topic. For every topic, we prepare a unified dataset that includes exclusive facts from three target languages (French, Russian, and Chinese). These facts are grouped by language code (\texttt{fr}, \texttt{zh}, \texttt{ru}) to support language-specific interaction features such as language filtering (D4). To reduce cognitive overload, we limit the number of displayed facts to \textbf{10} per language, proportionally sampling based on section-level gap counts. If fewer than 10 gaps exist for a language, we display all available facts.

\subsubsection{Translations and traceability.}
Because \textsc{WikiGap} displays cross-lingual content within the English Wikipedia page to readers who we do not assume read other languages, we translate the gap facts from their original language ($L_t$) into English for presentation in the fact cards. To support source traceability and deeper exploration, we create a direct \emph{link-to-highlight} for each fact. An encoded version of the original sentence in $L_t$ is appended to the target article's URL, sending readers to the exact sentence in context when they click on the ``View on [$L_t$] Wikipedia'' in the card (D3).

\subsubsection{In-text highlights.}  
\begin{figure}[htbp]
  \centering
  \includegraphics[width=\linewidth]{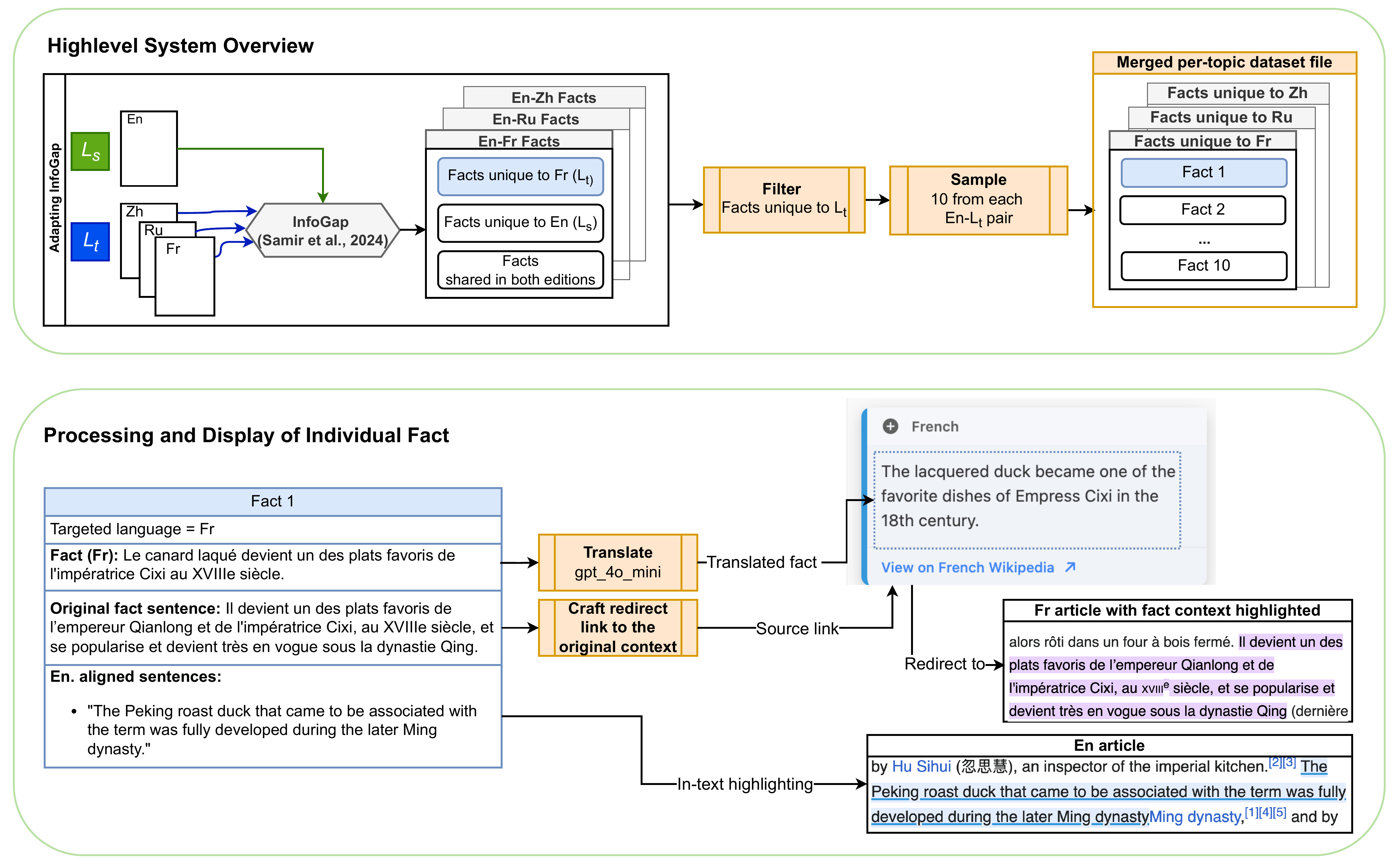}
  \caption{
    \textbf{Overview of system implementation and data processing pipeline.} 
    \textit{Top:} A high-level overview of the data stream in the \textsc{WikiGap} system. We adapted the \textsc{InfoGap} pipeline to support Chinese-language input alongside existing language pairs, followed by post-processing steps to standardize and merge datasets by topic. Orange process blocks indicate components we developed to enable proper integration and display of multilingual facts in the UI. 
    \textit{Bottom:} The data structure and rendering flow for an individual fact. This illustrates how each multilingual fact is transformed -- through translation, alignment, tagging, and contextual linking -- into an interactive component in the \textsc{WikiGap} interface.
  }
  \label{fig:infogap-ui-mapping}
\end{figure}

For each gap fact, \textsc{InfoGap} outputs the most semantically related English sentence from the source article, even when the fact itself is absent in English. We use these sentences as anchors for subtle in-text highlights (D1), which allow readers to preview cross-lingual differences inline without disrupting their reading flow. Figure~\ref{fig:infogap-ui-mapping} illustrates how such outputs map onto \textsc{WikiGap}'s interface components. For example, the French Wikipedia article for \emph{Peking Duck} mentions that the dish became a favorite of Empress Cixi in the 18\textsuperscript{th} century -- a detail missing from the English article. \textsc{InfoGap} aligns this fact to the closest English sentence ``The Peking roast duck that came to be associated with the term was fully developed during the later Ming dynasty'', 
%While this sentence does not convey the Empress Cixi detail, it provides a 
which provides a semantically-relevant anchor for surfacing the French-exclusive fact in context. 

\subsection{System Implementation}
\label{subsec:implementation}

We implemented \textsc{WikiGap} as a browser-based system that overlays multilingual content directly onto English Wikipedia articles. The system interface was developed as a Chrome extension using standard web technologies -- HTML, CSS, and vanilla JavaScript -- while the underlying data pipeline generates and serves precomputed JSON files derived from the \textsc{InfoGap} framework. When running \textsc{InfoGap}, we used GPT-4o for fact decomposition and fact alignment verification, and computed sentence embeddings using the multilingual LaBSE model \cite{openai2024gpt4o, feng2020language}. For each English Wikipedia topic, \textsc{InfoGap} produces three separate datasets, each containing factual gaps between English and one of three target languages (\texttt{fr}, \texttt{zh}, \texttt{ru}). We then merge these files by topic into a unified, standardized JSON file named after the English article. This consolidation supports runtime efficiency: when a user visits an English topic page, the extension loads the corresponding file based on the topic title. Each file contains multilingual facts translated into English which are generated using GPT-4o-mini \cite{openai_gpt4omini_2024}. The overall system workflow is illustrated in Figure~\ref{fig:infogap-ui-mapping}.

The extension interacts with the live English Wikipedia page, dynamically injecting highlights and sidebar content based on the current topic the user is browsing (the corresponding dataset is selected for rendering the data on the UI based on that topic). For each fact card, we generate external links that redirect readers to the original sentence in the source-language Wikipedia article. This is done by appending an encoded version of the original sentence to the article's base URL, allowing users to jump directly to the fact in context. Additionally, the search function is implemented by scanning both the fact body and language label for matches with the user's input and dynamically hiding non-matching entries.

% \begin{table}[t]
%   \centering
%   \small
%   \caption{Mapping \textsc{InfoGap} outputs to \textsc{WikiGap}'s dataset and interface components.}
%   \label{tab:InfoGap-mapping}
%   \begin{tabular}{p{3.8cm} p{5.2cm} p{4.0cm}}
%     \toprule
%     \textbf{\textsc{InfoGap} Output} & \textbf{Mapped Dataset Field} & \textbf{\textsc{WikiGap} UI Component} \\
%     \midrule
%     Separate dataset per target language per topic  
%       & Merged into a single topic file grouped by language  
%       & Language filter (D3) \\
%     \addlinespace
%     Original fact in the target language  
%       & Stored with English translation (via GPT-4o-mini)  
%       & Fact card content (D4) \\
%     \addlinespace
%     Source sentence (in target-language article)  
%       & Encoded as \textit{link-to-highlight} URL  
%       & “View on [$L_t$] Wikipedia” button (D4) \\
%     \addlinespace
%     Closest aligned English sentence  
%       & Used as anchor reference  
%       & In-text hover highlight (D1) \\
%     \bottomrule
%   \end{tabular}
% \end{table}

%% file: revised-sections/3-subsec-design-revised.tex
\subsection{Designing \textsc{WikiGap}}
\label{sec:design-ui}

We describe the exploratory interview we conducted to understand users' needs, from which we developed the design requirements (\S\ref{sec:design:reqs}). Then, we describe the core UI elements that we derived from the requirements (\S\ref{sec:design:ui}).

\subsubsection{Design Requirements Informed by User Needs and Theory}
\label{sec:design:reqs}
\begin{figure}[htbp]
\centering
\includegraphics[width=0.95\linewidth]{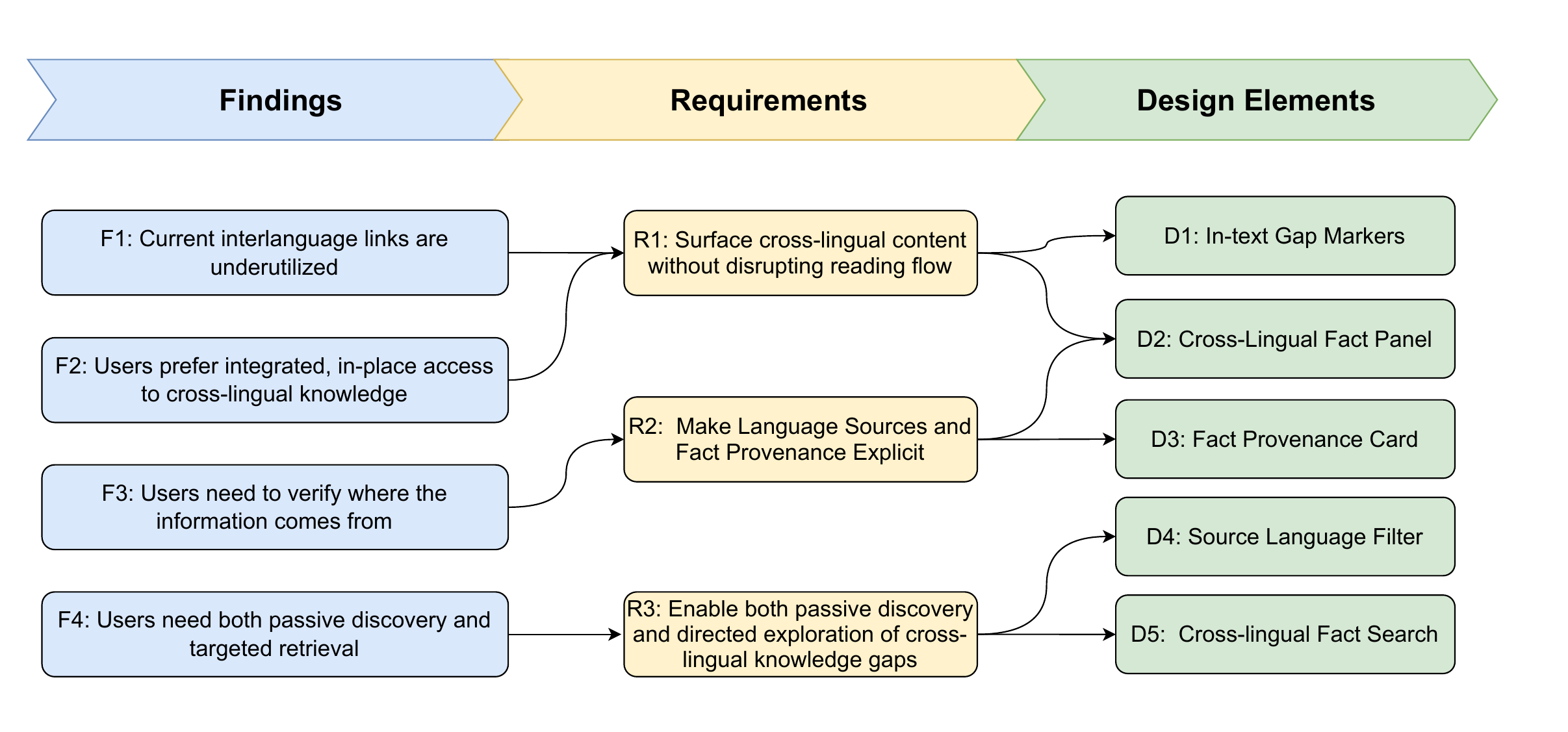}
\caption{How the findings from our preliminary interviews (F1-F4) informed \textsc{WikiGap}'s design requirements (R1-R3) and core design elements (D1-D5).}
\label{fig:findings_to_design}
\end{figure}

Guided by an asset-based perspective that shifts the focus from correcting gaps to making differences visible and interpretable, we combined findings from our exploratory interviews with established theories of reading and information seeking to define the design requirements for \textsc{WikiGap}. These requirements share an overarching goal: enabling readers to access cross-lingual knowledge while preserving the familiar experience of browsing English Wikipedia. The findings from the exploratory interview, requirements, and design elements  are summarized in Figure~\ref{fig:findings_to_design} and detailed below. %The \textsc{WikiGap} interface is illustrated in Figure~\ref{fig:ui}. 

\paragraph{Exploratory Interviews on Multilingual Wikipedia Use}

We conducted exploratory interviews with four Computer Science graduate students (2 women, 2 men), all regular Wikipedia users; three spoke an additional language (Korean, Russian, or Hindi). Each 30-minute session followed an unstructured, conversational format guided by open-ended questions about their multilingual Wikipedia usage, perceptions of the Wikipedia ILL system, and ideas for improving cross-lingual content presentation. Participants reviewed an English Wikipedia article on \emph{mooncake}--a topic unfamiliar to them--and reflected on how multilingual facts could be surfaced in this context. They proposed interface features to better support multilingual reading habits. We took notes during the sessions and synthesized them using thematic coding to derive the following design requirements for \textsc{WikiGap}.

% \begin{figure}[t]
%   \centering\includegraphics[width=\linewidth]{figures/Fig1.pdf}
%     \caption{The \textsc{WikiGap} interface embeds cross-lingual facts into English Wikipedia via five key design elements (D1–D5), supporting in-place access, traceability, and multilingual engagement.}
%   \label{fig:ui}
% \end{figure}

\paragraph{\textbf{R1: Surface cross-lingual content without disrupting reading flow.}}

This requirement directly addresses participants’ tendency to overlook multilingual content due to the high cost of switching language editions. Participants expressed a strong desire to access multilingual information without breaking their reading experience. They reported rarely using Wikipedia's ILLs (\textbf{F1}), noting the disorienting nature of full-page switches: \textit{``Even if I click the other language, I don't know where to look for the thing I care about. It's like starting over.''} Such page jumps impose high navigation costs \citep{pirolli1999information}, requiring users to reorient to a new layout and language, which discourages exploration.

Instead, users preferred lightweight, non-intrusive cues -- similar to annotations in Grammarly or Google Docs -- that subtly signal the presence of additional information (\textbf{F2}). This aligns with Anchored Annotation Theory \citep{marshall1997annotation}, which finds that tying annotations to specific text locations improves comprehension and recall. Our design anchors cross-lingual facts directly to the English article using click-triggered underlines and a collapsible panel, allowing users to access additional facts only when interested. This approach follows the Spatial Contiguity Principle~\citep{johnson2012eye}, which emphasizes placing related information close together. Collectively, these design choices preserve reading flow while enabling low-friction, in-place access to multilingual content.

\paragraph{\textbf{R2: Ensure clear language attribution and traceability of multilingual facts to support trust and credibility.}} This requirement reflects readers’ need to assess the provenance of multilingual facts, especially when such information is absent from the English article. In our interviews, a recurring concern among participants was the trustworthiness of information surfaced from other language editions--particularly when such facts were not present in the English version. Participants emphasized that they wanted to know \textit{``where the information comes from,``} including which language edition it originated from and the ability to read it in its original context (\textbf{F3}), because it is crucial for accessing information's credibility. This requirement highlights the importance of traceability and attribution in multilingual information access. Readers must be able to understand both the linguistic source and the original context of each surfaced fact. This entails more than just showing a translation, but rather also requires visual and interactive mechanisms that reinforce provenance and credibility.

To meet this need, we present each fact with an explicit, color-coded language label and allow readers to trace each fact back to its source by opening the corresponding language edition and viewing the sentence in context. These strategies support transparency and accountability, giving readers confidence in the surfaced content and encouraging deeper multilingual exploration.

\paragraph{\textbf{R3: Support both passive discovery and active exploration of cross-lingual knowledge gaps  (customizable exploration).}}

This requirement reflects the need to support different levels of engagement, recognizing that readers vary in how much cross-lingual information they want to encounter. User feedback revealed a range of interaction preferences: users want flexible access modes--both passive discovery and targeted retrieval (\textbf{F4}). Different participants had different expectations for how cross-lingual facts should be surfaced. Some favored passive discovery, where the system automatically highlights interesting differences without requiring user input. Others preferred more active control, such as being able to search for specific content or filter information by language or topic. They discussed concern over potential information overload if too many facts were presented at once. 

This requirement reflects that readers should be able to modulate how much cross-lingual content they see and which languages they want to include in their exploration. From a cognitive standpoint, this helps reduce information overload and supports a more goal-directed behavior. It necessitates features like search, filtering, and toggling between languages -- functions that empower users without requiring them to dig through entire articles in other languages.

% \begin{table}[t]
% \centering
% \caption{Design comparison between Wikipedia's Inter-Language Links (ILL) and \textsc{WikiGap}.}
% \label{tab:contrast_ill}
% \begin{tabular}{p{3.2cm}p{4.5cm}p{4.5cm}}
% \toprule
% \textbf{Design Dimension} & \textbf{ILL} & \textbf{\textsc{WikiGap}} \\
% \midrule
% Navigation cost & Full-page reload; reader loses position & Hover or click inline; context preserved \\
% Spatial contiguity & None (content shown on a separate page) & Anchored underlines with in-situ sidebar \\
% Language attribution & Compact link list; language implicit & Color-coded badges with explicit language labels \\
% Progressive disclosure & All content shown at once & Fact card $\rightarrow$ sidebar $\rightarrow$ source link \\
% Personalization & Not supported & Language filters; collapsible and pinnable sidebar \\
% Reading-flow disruption & High; interrupts user focus & Low; minimal visual disruption with subtle cues \\
% \bottomrule
% \end{tabular}
% \end{table}

\subsubsection{Core Design Elements}
\label{sec:design:ui}

Based on our design process and the identified requirements (R1-R3), we developed five core design elements for \textsc{WikiGap}, as illustrated in Figure~\ref{fig:ui}.

\paragraph{\textbf{D1: In-text Gap Markers.}} This design addresses R1 (non-disruptive presentation) and R2 (traceability) to make multilingual differences visible at the moment of reading without interrupting readers' flow, we implemented a subtle underlining system that signals the presence of additional information from other language editions. Each marker is country-color coded and appears as a patterned underline within the English article. When clicked, the side panel retrieves the fact relevant to that sentence, where the extension identifies complementary or missing information from another language edition. The underline then becomes more prominent, providing interaction feedback and inviting further exploration. Each language is associated with a distinct color -- red for Chinese, blue for French, and green for Russian. This color coding is applied across both in-text highlights and corresponding elements in the side panel, reinforcing source attribution as emphasized in R2. As we elaborate in Sec.~\ref{sec:wikigap:integration:language_fact}, only significant knowledge gaps identified by \textsc{InfoGap} are highlighted, rather than minor variations or differences in wording. This selective highlighting prevents information overload and ensures that users' attention is drawn to substantive cross-lingual differences.

\paragraph{\textbf{D2: Cross-Lingual Fact Panel}} To fulfill R1 and R2, we also developed a margin-anchored sidebar that displays the facts, translated to English and organized by their source language. When a user interacts with an in-text highlight, the corresponding fact card is revealed in the sidebar, which appears on the right side of the article. The sidebar is collapsible and can be pinned for extended browsing or hidden entirely to maintain a minimalist reading view. Each language section presents its facts in dedicated groups. Each fact is shown in English, accompanied by attribution information, including the source language tag and link to the original article. The close spatial relationship between in-text cues and sidebar content supports R1's emphasis on minimizing reading disruption.

\paragraph{\textbf{D3: Fact Provenance Card}} At the heart of \textsc{WikiGap}'s cross-lingual surfacing system is the fact card interface, which presents individual facts from other language editions in a compact, standardized format. Each fact card includes: (1) the target fact, translated into English, (2) a color-coded badge indicating the source language, and (3) a hyperlink labeled ``View on [Language] Wikipedia'' that opens the source article in a new tab with the original sentence highlighted. This design primarily addresses R2 (transparent attribution) by balancing at-a-glance comprehension with direct access to original context. 

\paragraph{\textbf{D4: Source Language Filter}} To support R3 (customizable exploration), we introduced a language filter function located at the top of the sidebar. These filters allow users to specify which language(s) they want to view content from, enabling personalization based on their linguistic background and interests. The filters use both language names and colors to maximize clarity. This functionality directly supports both passive and active interaction modes (R3). 
%The language filters are prominently displayed and use both language names and visual indicators (flags or icons) to ensure clarity. 
% The filtering system is particularly important for users who may be interested only in specific language editions or who want to reduce information overload by focusing on fewer languages at a time.

\paragraph{\textbf{D5: Cross-lingual Fact Search.}} To further support R3, specifically the directed exploration function, we added a search box feature that allows users to perform keyword-based queries across multilingual facts. When a user enters a search term, relevant fact cards from the selected languages are retrieved and displayed in the sidebar. This feature caters to users with goal-oriented information-seeking tasks, such as verifying a specific claim or comparing facts across language editions. While D1 and D2 emphasize passive and contextual discovery, the search box fulfills a complementary role by enabling intentional, focused retrieval. Together, these elements provide a spectrum of access modes that reflect users' varying preferences for how and when to engage with multilingual content.

% Together, these primitives...

% \textcolor{red}{[Add a concluding paragraph that restates the key insight from this section, e.g., ``\textsc{WikiGap} contributes novel UI primitives that address long-standing challenges in multilingual reading systems by...''. This helps the section land more like a contribution, not a spec sheet.]}

%% file: revised-sections/4-evaluation-revised.tex
% To validate our design, we conducted a mixed-methods user evaluation study, combining quantitative performance measures and qualitative feedback, which are summarized in Table~\ref{tab:study-metrics} and detailed below. We aim to understand \textsc{WikiGap}'s usability, its impact on knowledge gain, and user perceptions about the system and the cross‐lingual content.\footnote{The study protocol received approval from our institutional REB.} % ensuring that all participant data was collected and handled ethically.
To examine whether surfacing cross-lingual knowledge gaps within the English Wikipedia interface can meaningfully support multilingual engagement, we conducted a mixed-methods user study combining quantitative performance measures with qualitative feedback. The study assesses usability, knowledge acquisition, and users’ perceptions of the system and the cross-lingual content it surfaces. Table~\ref{tab:study-metrics} summarizes the study metrics and analysis.\footnote{The study protocol received approval from our institutional REB.}

\input{tables/metrics}

\subsection{Participants}
Twenty-one people participated in the study. Of these, eight were student volunteers from the HCI course where this project originated, and the remaining thirteen participants were recruited through Upwork, and were compensated at a rate of 20 USD for the user study. Participants ranged in age from 18 to 44 and represented diverse cultural and ethnic backgrounds (see Table~\ref{tab:participant-ethnicity}). All participants reported that they typically use the English version of Wikipedia. Twelve had previously contributed to Wikipedia to varying degrees, with two considering themselves frequent editors. Participants were required to meet the following inclusion criteria: (i) regular use of English Wikipedia, (ii) comfort with using a Chrome browser, and (iii) willingness to contribute user experience feedback. 

The study followed a within-subject design, in which each participant was assigned two different topics out of the total five topics (Sec.~\ref{subsection:topics}). Each participant read one article using the \textsc{WikiGap} extension and another article using the default Wikipedia interface (with ILLs) as a control. We selected interlanguage links (ILLs) as the control to reflect the default, in-context Wikipedia experience and to simulate an extension built on top of existing knowledge-seeking workflows; moreover, prior multilingual interface designs are not publicly available for deployment. The assignment of treatment and control conditions was randomized for each participant.

% \begin{figure}[t]
%   \centering\includegraphics[width=0.5\columnwidth]{figures/participants.pdf}
%   \caption{Demographics of Participants.}
%   \label{fig:participants}
% \end{figure}

\begin{table}[t]
  \centering
  \caption{Participant self-reported cultural background.}
  \label{tab:participant-ethnicity}
  \begin{tabular}{@{}l r@{}}
    \hline
    Cultural / Ethnic background & \% of participants \\
    \hline
    East Asian & 24\% \\
    White or European & 19\% \\
    South Asian & 14\% \\
    Hispanic or Latin American & 14\% \\
    Black or African & 10\% \\
    Middle Eastern or North 
    Southeast Asian & 10\% \\
    African & 5\% \\
    Other & 5\% \\
    \hline
  \end{tabular}
\end{table}

\subsection{Tasks}
Participants completed an open-book reading quiz consisting of 10 multiple-choice questions in each condition. For each quiz, participants were provided with four Wikipedia articles on the same topic—specifically, the English article and its corresponding versions in French, Russian, and Chinese. They were instructed to rely solely on these four articles to answer the quiz questions and were not allowed to use external resources such as Google Search. Each participant completed the quiz twice: once using the standard Wikipedia interface (control condition) and once using the \textsc{WikiGap} extension (treatment condition). For every question, participants were also asked to indicate the language edition in which they found the answer.

\paragraph{Quiz Construction.}
\label{sec:quiz-construction}
We constructed one quiz per article topic using facts extracted by \textsc{InfoGap} and surfaced through the \textsc{WikiGap} interface. From the 30 multilingual facts identified for each article, we manually selected a subset of 10, ensuring a roughly equal number of facts from each of the three language editions—French (fr), Russian (ru), and Chinese (zh)—to maintain balanced representation. For a fair comparison, we ensured that all quiz questions are answerable from InfoGap facts and by extension from the respective articles in French, Russian, and Chinese.

We then prompted a large language model GPT-4o~\cite{openai2024gpt4o} to generate a multiple-choice question with four answer choices from each fact. The questions were manually reviewed and edited to ensure clarity, factual correctness, and alignment with the presented content. In addition to selecting the correct answer, participants were asked to indicate the language edition in which they found the supporting fact. Although all quiz facts were sourced from non-English Wikipedia editions, we note that some participants may have inferred the correct answer from contextual clues in the English article. Nonetheless, the quizzes were designed to evaluate whether participants engaged with multilingual content. If participants had guessed entirely at random, the expected accuracy would be 25\%, given the four-choice format. Example questions are included in the Appendix~\ref{sec:sample_quiz_questions}.

\paragraph{Performance Metrics.} 
We recorded each participant’s task completion time and quiz accuracy score under both conditions. Quiz accuracy captured how much multilingual knowledge participants were able to obtain, while completion time reflected how quickly they could locate and interpret the relevant information. Taken together, these two measures provide an indication of learning efficiency and allowed us to quantitatively compare the effectiveness of \textsc{WikiGap} with the standard Wikipedia interface.

%The mixed-methods metrics of the study and the respective data collection and analysis are described in Table~\ref{tab:study-metrics}.

\subsection{Materials and Topic Assignment}
\label{subsection:topics}

\paragraph{Selected topics}
% We are interested in looking at food articles in geneal as food discourse provides a particularly telling lens through which to study knowledge gaps, because it is deeply rooted in regional traditions and cultural practices \cite{Luo2023OtheringAL}.
% We selected five dishes, each anchored in a distinct national cuisine—\emph{Injera} (Ethiopia), \emph{Paella} (Spain), \emph{Philippine Adobo} (Philippines), \emph{Peking Duck} (China), and \emph{Wiener Schnitzel} (Austria)—as the article topics shown in the \textsc{WikiGap} extension.
We focus on food articles because food serves as a culturally rich lens through which to explore knowledge gaps across language editions in Wikipedia. Culinary practices are deeply embedded in regional customs, social histories, and national identities, making food a meaningful proxy for cultural knowledge \cite{Luo2023OtheringAL, winata-etal-2025-worldcuisines}. Analyzing food allows researchers to capture both shared and divergent cultural representations, such as differences in dish names, preparation styles, and regional associations \cite{winata-etal-2025-worldcuisines}. This cultural complexity makes food an ideal domain for surfacing factual asymmetries. We selected five culturally specific dishes: \emph{Injera} (Ethiopia), \emph{Paella} (Spain), \emph{Philippine Adobo} (Philippines), \emph{Peking Duck} (China), and \emph{Wiener Schnitzel} (Austria). Each of these dishes has a dedicated Wikipedia article and distinct cultural heritage. % as the focus topics shown in the \textsc{WikiGap} extension. 
Table~\ref{tab:knowledge_gap_summary} summarizes the number of language-exclusive facts detected per topic and the total number shown in the \textsc{WikiGap} interface.

All five dishes satisfy the following practical constraints: (i) they originate outside the English-speaking world, providing a comparable degree of cultural distance from the English source language, (ii) their English Wikipedia article is moderately sized (1,000–2,000 words), and (iii) the \textsc{WikiGap} extension can surface \textasciitilde30 multilingual facts for each.\footnote{\emph{Oolong tea} was included in the pilot with the first two participants but was dropped because its cultural context overlapped substantially with \emph{Peking Duck}. It was therefore replaced with \emph{Philippine Adobo} to preserve cultural diversity in topic representation.}

\input{tables/topics}

% \paragraph{Pre‑computing topic datasets.} To ensure a responsive and scalable experience, we pre‑compute the output of \textsc{InfoGap} for each topic rather than running the pipeline in real time. 

\paragraph{Topic assignment}
Each participant is assigned two distinct topics, one for each condition (control and \textsc{WikiGap}). Topic assignment was designed to minimize cultural confounds and familiarity‑driven performance differences. Specifically, the food topic chosen for each participant could not be strongly linked to the participant's self‑reported cultural or ethnic background. To reduce individual-performance bias, each topic was assigned in every condition to \textbf{exactly four} different participants.

\subsection{Procedure}
The lead investigator conducted the study remotely over a 1–1.5 hour recorded Zoom call. At the start, participants received a zip file containing the \textsc{WikiGap} extension and were instructed to install it on their Google Chrome browser. A short tutorial was provided by the researcher for both the treatment and control conditions. During the treatment condition, we began the round by introducing the \textsc{WikiGap}'s functionality and allowed participants to explore the extension on their own until they felt ready to proceed. In the control condition, we demonstrated how to access language versions on Wikipedia using ILLs, and how to use Chrome's Google Translate function to read the content in English. At the beginning of each round, one of the five topics was assigned to the participant according to the topic assignment rules. After completing each round, the participant completed a System Usability Scale (SUS) after finishing the quiz. We used the SUS to measure ease-of-use and users' preference on with or without the \textsc{WikiGap} extension. To conclude, the first twelve participants completed an open-ended questionnaire, while the remaining nine took part in a semi-structured interview. Both instruments covered the same six areas: overall impressions of the system, presentation of multilingual information, perceived cultural impact, trust in the augmented facts, perceptions of Wikipedia’s reliability after using \textsc{WikiGap}, and future inclination to explore multilingual content (see Appendix~\ref{sec:protocol} for the full list of questions).

% \subsubsection{Analysis}
% The mixed-methods metrics of the study and the respective data collection and analysis are described in Table~\ref{tab:study-metrics}.

%% file: tables/metrics.tex
\begin{table}[t]
\centering
\small
\setlength{\tabcolsep}{3pt}
\caption{\tocheck{Description of study metrics and corresponding data collection and analysis. 
All metrics were collected at the individual participant level and compared across \textsc{WikiGap} and control conditions.}}
\label{tab:study-metrics}
\begin{tabular}{lp{5.7cm}p{6.7cm}}
\toprule
\textbf{Metrics} & \textbf{Data Collected} & \textbf{Data Analysis} \\
\midrule
\textbf{Performance} & & \\
Quiz Accuracy & Percentage of correct answers in each condition & \multirow{2}{*}{Paired t-test comparing \textsc{WikiGap} vs. control conditions} \\
Completion Time & Time (in minutes) to complete each quiz session &  \\
\midrule
\textbf{Preference} & & \\
Usability & System Usability Scale (SUS) score in each condition & Paired t-test comparing SUS scores between conditions \\
User Feedback & Open-ended responses and interview transcripts & Thematic coding to extract perceived usability, utility, and challenges \\
\bottomrule
\end{tabular}
\end{table}

%% file: tables/topics.tex
\begin{table}[t]
\centering
\caption{Number of knowledge gap facts discovered per food topic and language by \textsc{InfoGap}, and the total number of facts shown in the WikiGap interface. For each topic, up to 10 facts per language were selected to ensure balanced representation and reduce cognitive overload. \textit{Injera} had only 8 gaps in Chinese, resulting in 28 total facts shown (*).}
\begin{tabular}{lcccc}
\hline
\textbf{Food Topic} & \textbf{Russian (ru)} & \textbf{French (fr)} & \textbf{Chinese (zh)} & \textbf{Facts Shown in WikiGap} \\
\hline
Wiener schnitzel    & 65 & 20 & 62 & 30 \\
Peking duck         & 13 & 23 & 69 & 30 \\
Paella              & 28 & 62 & 52 & 30 \\
Philippine adobo    & 21 & 15 & 84 & 30 \\
Injera              & 10 & 14 & 8  & 28* \\
\hline
\end{tabular}
\label{tab:knowledge_gap_summary}
\end{table}

%% file: revised-sections/5-findings-revised.tex
We report findings from our mixed-method evaluation of \textsc{WikiGap}. Together, these findings show how making multilingual knowledge differences visible within the English Wikipedia interface affects users’ performance, preferences, and engagement with non-English content. Across quantitative measures and qualitative feedback, participants consistently described shifts in how they accessed information, perceived Wikipedia’s completeness, and engaged with multilingual knowledge during routine fact-finding.

\paragraph{\textbf{1. Making Multilingual Differences Visible Improves User Preference and Reading Experience (D1, D2, D3)}}

We evaluated user preference using the System Usability Scale (SUS). The average SUS scores for each condition are presented in Figure~\ref{fig:sus}, along with general SUS cutoff scores for fair, good, and excellent usability \citep{sus}. Overall, \textsc{WikiGap} achieved a substantially higher usability score compared to the control. \textsc{WikiGap} falls within the ``excellent'' usability category, while the default Wikipedia page with ILL scored below the ``fair'' threshold. A paired-samples \textit{t}-test revealed a significant difference in scores between the \textsc{WikiGap} condition (\textit{M} = 82.5, \textit{SD} = 10.9) and the control condition (\textit{M} = 48.2, \textit{SD} = 18.6), \textit{t}(20) = 6.60, \textit{p} < .001. (Figure~\ref{fig:sus}). 

\definecolor{burntorange}{rgb}{0.8, 0.33, 0.0}
\definecolor{darkcerulean}{rgb}{0.03, 0.27, 0.49}

\begin{figure}[t]
  \centering
  \includegraphics[width=0.5\linewidth]{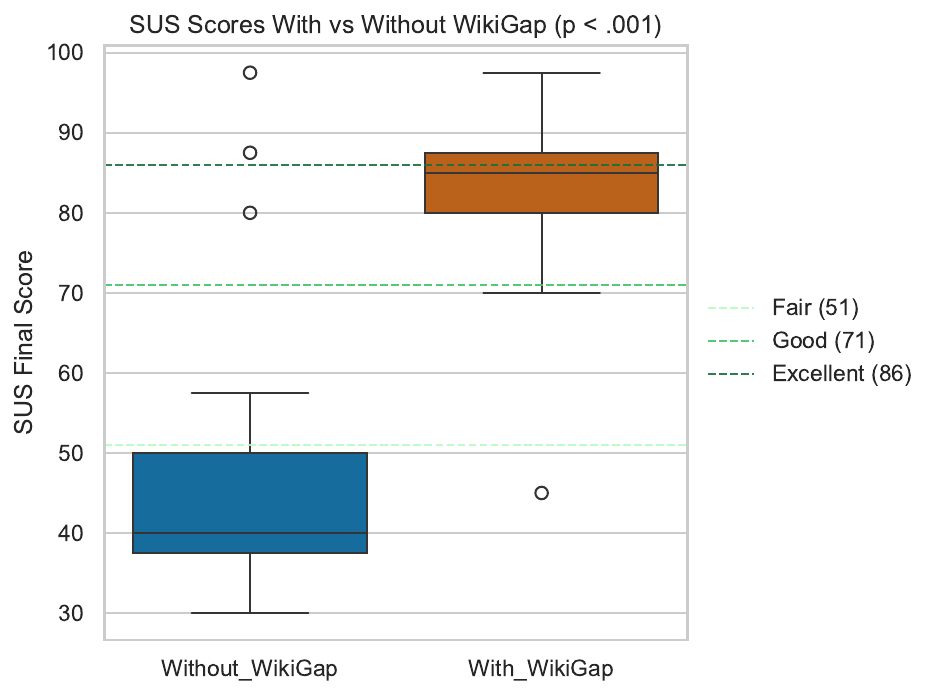}
  \caption{
  Box plot showing the System Usability Scale (SUS) scores for each condition. \textcolor{darkcerulean}{Blue} represents the control condition (no \textsc{WikiGap}), and \textcolor{burntorange}{orange} represents the treatment condition (with \textsc{WikiGap}). Horizontal dashed lines represent standard usability benchmarks in varying shades of green: \textcolor[HTML]{7CCB8A}{light green} for \textit{Fair} usability (SUS > 51), \textcolor[HTML]{40C463}{medium green} for \textit{Good} usability (SUS > 71), and \textcolor[HTML]{196F3D}{dark green} for \textit{Excellent} usability (SUS > 86).
  }
  \label{fig:sus}
\end{figure}

Participants attributed their strong preference for \textsc{WikiGap} not only to its ability to surface knowledge disparities, but also to the various design elements (D1-D3) that supported the quick access and ease of reading. Many highlighted how having access to multilingual content directly within the English article -- through translated facts in a sidebar (D2, D3) -- lowered the barrier to engaging with other language editions. It helped users ``absorb information without the language difference obstacle'' (P8). Participants appreciated the sidebar's structure and organization by language, which kept all cross-lingual facts ``apparent within a single page'' (P5), eliminating the need to navigate away and reducing friction in multilingual exploration. In contrast to ILL's full-page switch model, participants noted that the \textbf{sentence-level fact card interface (D3)} made the information easier to process. As P11 remarked, \textit{``[WikiGap] presents sentences instead of long paragraphs from the Wikipedia page.''}

Other visual elements also shaped participants' perceptions of usability. Several users found the \textbf{in-text gap markers (D1)} helpful in surfacing areas of missing information in the English article. However, reactions to the color scheme used for language cues were mixed. While some found the colors intuitive, others, like P19, felt that the red-green coding could unintentionally suggest \textit{``right or wrong''}.

Despite minor confusion about the color scheme, most users described the overall experience as more enjoyable and cognitively less demanding than their usual experience with Wikipedia's default ILL interface. P9 summarized this sentiment: \textit{``Sometimes even when I open my Wikipedia, I would not read that much because of laziness, but \textsc{WikiGap} did a good job extracting information and makes the whole experience much more pleasant.''}

\paragraph{\textbf{2. Visibility of Cross-Lingual Differences Improves Fact-Finding Performance and Supports Cross-Cultural Learning (D4, D5)}}

% \paragraph{\textbf{2. \textsc{WikiGap} Enhanced Fact-Finding Performance and Cross-Cultural Learning via Passive Discovery and Active Exploration (D4, D5)}}

\begin{figure}[t]
    \centering
    \includegraphics[width=.9\linewidth]{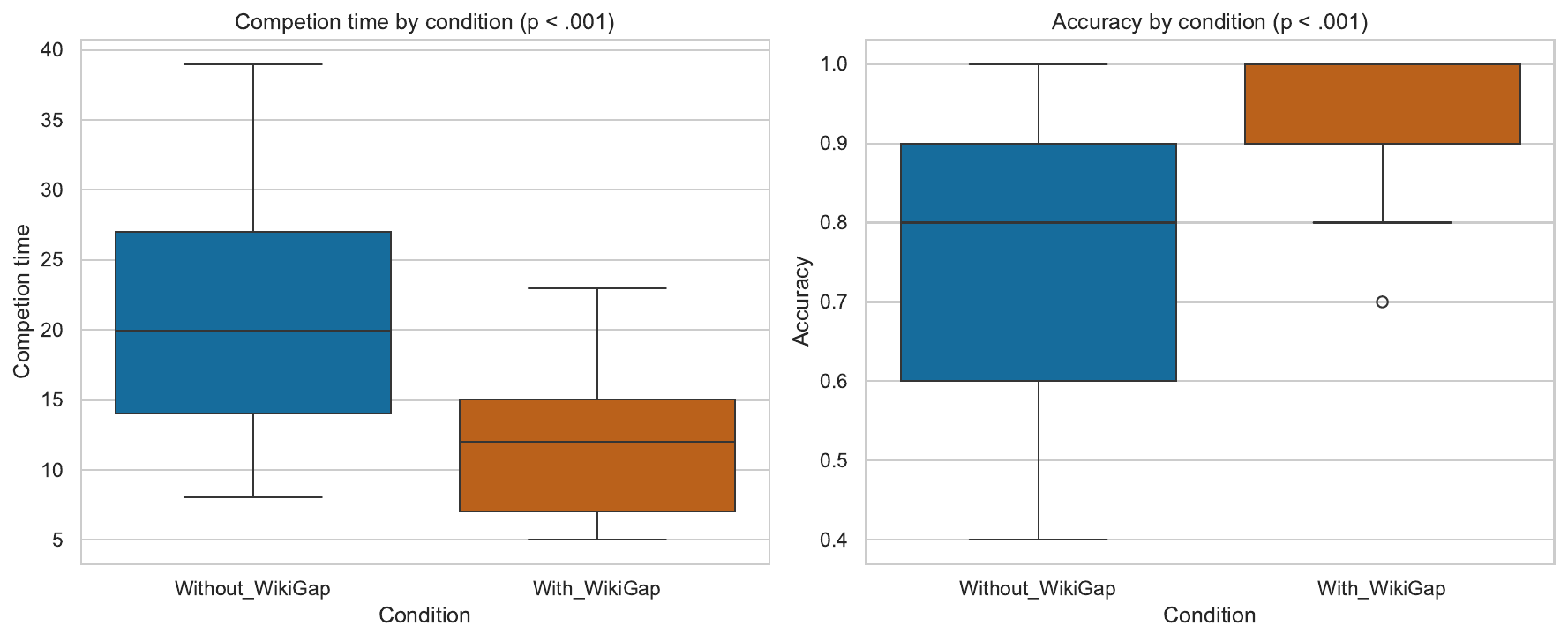}
    \caption{
       \textbf{Average completion time} (left) and \textbf{quiz accuracy} (right) across conditions. \textcolor{darkcerulean}{Blue} represents the control condition (no \textsc{WikiGap}), and \textcolor{burntorange}{orange} represents the treatment condition (with \textsc{WikiGap}). Using \textsc{WikiGap} demonstrates a statistically significant improvement in user performance on the quiz.
    }
    \label{fig:quiz}
\end{figure}

To evaluate the impact of \textsc{WikiGap} on users' ability to retrieve and retain cross-lingual information, we measured task performance through a fact-finding quiz, using both accuracy and completion time as metrics. As shown in Figure~\ref{fig:quiz}, participants performed significantly better when using \textsc{WikiGap}, achieving higher quiz accuracy and completing the task more quickly than with the default Wikipedia interface. A paired-samples \textit{t}-test showed that accuracy was significantly higher in the \textsc{WikiGap} condition ($M$ = 0.91, $SD$ = 0.09) than in the control condition ($M$ = 0.73, $SD$ = 0.19), \textit{t}(20) = 4.75, \textit{p} < .001. Completion time also improved significantly, with users finishing the quiz faster using \textsc{WikiGap} ($M$ = 12.39 minutes, $SD$ = 5.78) compared to the control ($M$ = 20.84 minutes, $SD$ = 8.68), \textit{t}(20) = -5.52, \textit{p} < .001.\footnote{Faster completion times in the \textsc{WikiGap} condition is partly expected, as the quiz questions were constructed from facts extracted by \textsc{InfoGap}. In real-world use, not all facts would be presented by \textsc{WikiGap}, and completion time gains may therefore be more modest.} These results suggest that \textsc{WikiGap} not only facilitated access to relevant information but also improved users' efficiency in locating and interpreting cross-lingual facts. \textsc{WikiGap} consistently improved user performance across all five culturally diverse topics. For a topic-by-topic breakdown of quiz accuracy and completion time, see Appendix Figure~\ref{fig:quiz_detailed}.

These performance gains were reinforced by participants' qualitative feedback, which highlighted the value of both passive discovery and active exploration features. \textbf{The fact search (D5)} enabled users to directly retrieve relevant facts, supporting goal-oriented behaviors and speeding up information retrieval. As P16 explained, \textit{``I don't need to go to multiple pages. I can search in the \textsc{WikiGap} extension and find any information quickly. Using the \textsc{WikiGap} extension saved me a lot of time.''} This targeted retrieval helped users answer questions more efficiently, contributing to the improved fact-finding performance observed in the quiz metrics. Complementing this, the \textbf{language filter controls (D4)} allowed users to personalize their exploration by prioritizing facts from languages culturally aligned with the topic. Several participants reported that they first filtered to the language they deemed most relevant, as they\textit{``will trust more when the language is directly aligned with the original country of the food.''} (P1). Similarly, P7 remarked, \textit{``I would love to see how the Tagalog page talks about Philippine adobo. In fact, diverse cultural perspectives on food is more linked to the cultural ecosystem where the food is grounded.''}. Together, D4 and D5 supported efficient, goal-driven access to cross-lingual content.

Beyond performance gains, the extension also facilitated passive discovery and cross-cultural learning -- a central goal of our design. Several users remarked that they encountered information they would not have actively sought out on their own. As P1 reflected, \textit{``Without [\textsc{WikiGap}], I wouldn't have thought about looking into Wikipedia pages in different languages.''} Others described moments of unexpected cultural insight, such as P5, who noted, \textit{``I would have completely missed out on the historical aspect if I used Wikipedia normally (only in English).''} Participants also pointed to surprising culture-specific facts, like the culinary use of Sprite in a traditional dish (P13). These examples illustrate that making multilingual differences visible not only improved task efficiency, but also broadened users’ exposure to culturally situated knowledge.

% To assess the appropriateness of a paired-sample t-test, the Shapiro-Wilk test was conducted to verify the normality of the difference scores between the \textsc{WikiGap} and control conditions. For both metrics, the test failed to reject the null hypothesis of normality (accuracy: p = 0.5086; completion time: p = 0.2529), confirming the use of a paired t-test. 

\paragraph{\textbf{3. \textsc{WikiGap} Raises Awareness of Multilingual Gaps and Motivates Exploration}}
\label{sec:results:qualitative:aware}

Participants consistently credited \textsc{WikiGap} for revealing factual disparities they were previously unaware of. Several users described being surprised that different language versions did not simply mirror each other. As P20 reflected, ``\textit{Before \textsc{WikiGap}, I always assumed that everything presented in English would also be visible in another language.''} Similarly, P18 noted that they ``\textit{didn't know there could be so many differences in how the content was presented in various languages.''}

This realization often prompted frustration as participants became aware of the limits of English Wikipedia. As P5 expressed, \textit{``I'm frustrated that this doesn't already exist! I feel like I've been missing out on information from other languages because I just assumed other pages would have the same information translated.''} Other participants had deeper reflections on how \textsc{WikiGap} changed their perception about knowledge: \textit{``Knowledge is not plain, knowledge has different dimensions waiting to be discovered, and this tool offers this opportune moment.''} (P10).

Importantly, increased awareness also motivated future engagement. Participants described being more inclined to explore content outside their primary language now that differences were visible and easy to access. 
\begin{quote}

``\textit{I was never on the pages of Wikipedia where the language is not what I speak, because I often assume that all pages under different languages are the same if they are of the same topic, so sometimes I struggled to find some information in the Wikipedia. But if I have \textsc{WikiGap}, I think I will definitely explore more.''} — P9 
\end{quote}

Several emphasized that centralizing multilingual content within a single interface reduced the effort required to engage, making exploration more likely. Now they can \textit{``see different viewpoints from different language pages``} (P17), they are more inclined to explore content outside their primary language in the future. Several users described feeling more motivated to engage with multilingual content now that they understood its value. As P1 noted, ``\textit{If there's a tool like \textsc{WikiGap} that can help me easily get various information from other pages, I would like to explore. If not, I wouldn't navigate multiple pages myself.''}

\paragraph{\textbf{4. Respondents' Views on LLMs and Wikipedia: Perceived Trust and Tensions}}
\label{sec:results:qualitative:llm}
Awareness of cross-lingual disparities also led participants to reflect on the trustworthiness of Wikipedia and how it compares to LLM-based knowledge sources. We observed a wide range of perspectives. For some, \textsc{WikiGap} surfaced surprising inconsistencies between language editions. Those differences made them question the completeness or neutrality of English Wikipedia. Others interpreted these variations not as flaws, but as a feature of Wikipedia's multilingualism and editorial diversity. These divergent views led participants to compare Wikipedia to emerging alternatives such as LLMs, positioning \textsc{WikiGap} as a catalyst for rethinking how people evaluate and engage with knowledge sources.

A minority (n=2) of the participants reported a decline in trust in Wikipedia after using \textsc{WikiGap}. They described how uncovering missing or differing facts between language editions made them feel less confident in the platform's ability to represent comprehensive or balanced information. This, in turn, increased their inclination to rely on LLMs as knowledge sources:

\begin{quote}
\textit{``I feel that after using \textsc{WikiGap}, I would be inclined to use Wikipedia even less. Before this, I assumed that Wikipedia would be a reliable and complete source of knowledge. However, I now know that the English version misses several key details. This leads me to be more reliant on LLM-based tools that have all the information in one place.''} — P2
\end{quote}

For this participant and others, \textsc{WikiGap} altered their perception of Wikipedia as a trustworthy authority. The perceived comprehensiveness of LLMs, combined with their ability to synthesize content across sources, positioned them as a more efficient alternative, even if they lacked transparency.\footnote{In practice, the assumption that multilingual LLMs such as ChatGPT and Gemini can synthesize information learned from web text in different languages is not entirely true; they often fail to retrieve facts learned in one language when prompted in another language \cite{goldman2025eclekticnovelchallengeset}.}

However, not all participants responded this way. Others voiced strong concerns about the lack of attribution and provenance in LLM-generated responses, and emphasized that \textsc{WikiGap} made them appreciate Wikipedia's editorial structure and citation practices even more. For these users, the ability to trace facts back to their original source, including reading the sentence in the original language edition, was critical to their sense of trust:

\begin{quote}
\textit{``I don't trust ChatGPT that much because sometimes it just makes things up, and it doesn't provide sources. With \textsc{WikiGap}, I can click on, say, the French Wikipedia page, read it, and see the sources. It feels more reliable. I also appreciate that Wikipedia cites everything, so there's some quality control, even if it's not always perfect.''} — P13
\end{quote}

These contrasting responses underscore an important dynamic in how users engage with knowledge platforms: while LLMs offer convenience and perceived comprehensiveness, they can lack the transparency and editorial accountability that are central to Wikipedia's epistemic model. For some users, \textsc{WikiGap} undermined their confidence in Wikipedia. For others, it reaffirmed the value of its provenance-driven structure, even in the face of inconsistency. We return to the socio-technical implications in Sec.~\ref{section:discussion}.

\paragraph{\textbf{5. \textsc{WikiGap} Suggests Opportunities for Supporting Cross-Lingual Editing}}

Although \textsc{WikiGap} was primarily designed to support readers by surfacing factual cross-lingual content, a few participants, particularly those with prior editing experience, reflected on how it could assist with editorial tasks. These insights point to an emergent opportunity for \textsc{WikiGap} to support not only knowledge consumption but also knowledge curation on Wikipedia.

Some users noted that the ability to trace and verify information from other language editions made the tool useful for improving article quality. As one frequent editor explained, \textit{``Sometimes an article in another language has crucial information that's missing in English. \textsc{WikiGap} helps me find and translate it, then cite it properly.''} (P16). Another participant elaborated on how the tool fits into their existing editing workflow: \textit{``I'd check the source \textsc{WikiGap} shows me, then cite the same source... \textsc{WikiGap} helps me discover facts and references.''} (P17). While these reflections were not prompted by editor-specific questions, they point to an emergent opportunity: making multilingual gaps visible to readers may also lower barriers to cross-lingual knowledge integration, extending \textsc{WikiGap}’s impact beyond consumption to potential curation.

%% file: revised-sections/6-discussion-revised.tex
We interpret our findings through the lens of \textit{boundary objects}—shared artifacts that enable coordination across heterogeneous communities without requiring consensus or uniform interpretation \citep{star1989institutional,susan2010}. We argue that \textsc{WikiGap} functions as a boundary object within Wikipedia’s multilingual ecosystem by making cross-lingual knowledge differences legible, actionable, and traceable within everyday reading practices. In this role, \textsc{WikiGap} enables three intertwined forms of epistemic and coordinative work. First, it reframes readers’ epistemic assumptions by exposing the incompleteness of English Wikipedia and foregrounding knowledge produced by other language communities~(\S\ref{sec:dis:promote}). Second, it provides a shared reference point that supports articulation work between readers and editors across language editions, enabling verification and coordination without enforcing uniformity~(\S\ref{sec:dis:articulate}). Third, it gestures toward a broader infrastructural role by modeling how pluralistic, provenance-aware interfaces can counter knowledge consolidation in emerging AI-mediated systems~(\S\ref{sec:dis:llm}).

\subsection{Promoting Epistemic Equity by Challenging the ``English-as-Superset'' Assumption}
\label{sec:dis:promote}

Our user study revisits prior findings that the dominance of English Wikipedia leads readers to treat it as a comprehensive superset of knowledge across language editions, rather than as a culturally situated and socio-geographically positioned construction \citep{kumar2021digital}. Participants explicitly articulated this belief. Although more than 60\% of participants reported a non-English primary language, all identified English as their primary source on Wikipedia, largely because English Wikipedia covers more topics and is updated more frequently. Despite being familiar with Wikipedia's ILLs, participants predominantly consumed content in English, reflecting how infrastructural dominance concentrates reader attention within a single language edition and reinforces the invisibility of knowledge produced by other language communities. At the interface level, Wikipedia's current design renders other language editions technically accessible but epistemically invisible: ILLs signal the existence of alternative versions, yet do not communicate that English content is neither exhaustive nor neutral. This gap was evident in participants' assumptions; as one remarked, \textit{``Before WikiGap, I always assumed that everything presented in English would also be visible in another language.''}

Consistent with this perception, participants strongly preferred \textsc{WikiGap} over the default ILLs interface and performed significantly better on fact-finding tasks, indicating that access alone—without cues about epistemic differences, provenance or community context—is insufficient to support meaningful cross-lingual engagement. \textsc{WikiGap} challenges the ``English-as-superset'' assumption by both raising awareness of cross-linguistic knowledge gaps and redistributing reader attention toward knowledge produced by other language communities. Adopting an asset-based design orientation, the system foregrounds non-English content not as supplementary or corrective, but as a valuable product of distinct peer production efforts.

By embedding sentence-level complementary facts in context and pairing them with explicit provenance, \textsc{WikiGap} makes the limits of English Wikipedia visible and simultaneously inviting engagement with the work of other language editions. This design encourages readers to recognize Wikipedia as a collection of interconnected yet distinct knowledge production communities, rather than a single monolithic source of global consensus. In doing so, \textsc{WikiGap} promotes epistemic equity in a scoped, design-operational sense: by redistributing attention and engagement across Wikipedia’s multilingual peer production communities and supporting cross-community interaction as part of everyday readership. 

% This design encourages readers to recognize knowledge as partial, situated, and collectively produced, shifting attention away from an assumed global consensus toward the plurality of perspectives that constitute Wikipedia's multilingual ecosystem. In doing so, \textsc{WikiGap} promotes epistemic equity by inviting cross-lingual exploration and engagement as a natural part of everyday readership.

\subsection{WikiGap as a Boundary Object Supporting Articulation Work Across Reader and Editor Communities}
\label{sec:dis:articulate}
Wikipedia's multilingual knowledge production spans heterogeneous communities with distinct linguistic, cultural, and editorial norms. Our findings show that \textsc{WikiGap} operates as a boundary object in practice, providing a shared point of reference through which readers and editors from different language communities can interpret, evaluate, and act on cross-lingual knowledge differences without requiring consensus. By surfacing sentence-level complementary facts with clear provenance, \textsc{WikiGap} offers a way to support shared understanding across communities without requiring consensus or uniformity. Building on these findings, we interpret \textsc{WikiGap} as enabling an asset-based stance: it helps readers and multilingual contributors recognize non-English facts as usable resources by making them legible and actionable through provenance and in-context integration.

\subsubsection{For Readers of English Wikipedia.}
\textsc{WikiGap} benefits readers by supporting fact-finding, cross-cultural learning, and interpretive flexibility. Participants described discovering content they ``would have otherwise missed'' in English and expressed surprise at how much meaningful information existed exclusively in other language editions (Findings 1--2). Several noted that the tool increased their curiosity and willingness to visit other language pages, shifting their sense of Wikipedia from a monolithic source of global consensus to a collection of situated, culturally inflected knowledge communities. 

Through a CSCW lens, this suggests that \textsc{WikiGap} enhances the \textit{legibility} of multilingual knowledge work. It exposes facts that English readers typically never encounter, foregrounds the situated perspectives of marginalized linguistic communities, and supports interpretive flexibility---one of the core affordances of boundary objects. Rather than collapsing perspectives into a single authoritative narrative, \textsc{WikiGap} allows readers to compare, contextualize, and reinterpret information based on their own background knowledge and cultural standpoint.

\subsubsection{For (Potential) Editors Across Languages.}
Although our study focused on readers, participants with editing experience highlighted how \textsc{WikiGap} could support multilingual editorial workflows, they pointed to using the provenance link to see the original facts from the other languages to verify and potentially transfer content across editions. Maintaining consistency across language editions requires editors to reconcile differing source traditions, cultural emphases, and update rhythms. This process is inherently distributed and involves articulation work \citep{schmidt_taking_1992}: interpreting discrepancies, assessing their relevance to local communities, coordinating with other editors, and making decisions about whether and how to incorporate missing knowledge.

Participants noted that \textsc{WikiGap} reduces some of this articulation burden by automatically surfacing contextualized discrepancies, highlighting their provenance, and linking directly to original sentences. These cues help editors evaluate the origins and significance of differences, understand when they reflect meaningful cultural divergence versus simple incompleteness, and coordinate potential updates. 
Rather than automating agreement or enforcing uniformity, \textsc{WikiGap} supports the human work of interpretation and negotiation that underlies Wikipedia's collaborative ecosystem. By enabling multiple situated viewpoints to coexist while still facilitating coordination across them, the system exemplifies how boundary objects can scaffold the complex articulation work required for distributed knowledge production.

In this way, \textsc{WikiGap} renders the often invisible coordinative labor of multilingual maintenance more visible and manageable by making cross-lingual differences legible and actionable through provenance and in-context integration, reducing coordination friction and supporting editor engagement.

\subsection{Positionality, LLMs, and the Future Role of WikiGap as Boundary Infrastructure}
\label{sec:dis:llm}
Across both quantitative preferences and qualitative interviews, participants expressed strong appreciation for \textsc{WikiGap}'s explicit display of provenance and its framing of multilingual knowledge as situated rather than objective. 
% This preference for pluralism became particularly salient when contrasted---often spontaneously---against the increasingly common practice of using Large Language Models (LLMs) as consolidated knowledge sources.
This recognition of positionality became particularly salient when contrasted—often spontaneously—against the increasingly common practice of using Large Language Models (LLMs) as ostensibly objective, consolidated knowledge sources.

A small subset of participants (n=2) expressed a desire for consolidated knowledge and therefore perceived gaps highlighted by \textsc{WikiGap} as evidence of English Wikipedia's ``incompleteness.'' One participant noted that they would prefer to turn to LLMs, which present a single synthesized output. While LLM outputs are often perceived as authoritative or persuasive \citep{chockkalingam-etal-2025-go}, they rarely foreground multiple cultural or linguistic perspectives and frequently exhibit English-centric biases \citep{hershcovich-etal-2022-challenges,Mihalcea_Ignat_Bai_Borah_Chiruzzo_Jin_Kwizera_Nwatu_Poria_Solorio_2025, tao2024cultural}. They also obscure provenance \citep{shah2022situating}, making it difficult for users to evaluate where information originates. LLMs thus risk presenting a ``view from nowhere,''~\citep{haraway2013situated} even though their outputs reflect specific training data and design choices.

Although \textsc{WikiGap} and LLMs appear to sit at opposite ends of the spectrum---positionality versus objectivity—they need not be mutually exclusive. Our findings suggest that asset-based design principles foregrounding positionality can inform future LLM interfaces by encouraging users to see responses as situated rather than uniformly objective. Concretely, interfaces could surface multiple culturally grounded viewpoints instead of a single synthesized answer, along with lightweight provenance cues indicating which language communities or traditions each perspective reflects. Such designs would counter the tendency of current systems to amplify majority perspectives and obscure minority ones, helping users explore diverse interpretations more intentionally. They would also make latent cultural knowledge in LLMs more accessible to researchers and end users alike. In this sense, embedding \textsc{WikiGap}-like scaffolding into LLM interfaces represents a promising direction for foregrounding epistemic positionality within increasingly consolidated AI ecosystems. This extends our system’s asset-based design principles to cross-lingual and multilingual LLM interfaces, encouraging users to engage with knowledge as situated, grounded in provenance, and produced by distinct language communities.

\subsection{Limitations and Future Work}
\label{sec:limitations}
\paragraph{Language Coverage.}%
Our study focuses on English and three target languages (French, Russian, and Chinese). While these languages span different linguistic families and large Wikipedia communities, they do not capture the full diversity of Wikipedia's language ecosystem. Several participants expressed a desire for broader language coverage, particularly for pages in their language of origin. For example, one participant noted,  
\textit{``I would love to see how the Tagalog page could complement it. In fact, diverse cultural perspectives on food are more linked to the cultural ecosystem where the food is grounded''} (P7).   
Expanding to more languages will enrich cross-cultural perspectives and align with users' expectations in future work. 
The underlying \textsc{InfoGap} pipeline is language-agnostic: in principle, it can operate on \emph{any} language that GPT-4 supports, though quality may vary, especially for low-resource languages. This is important for future research as smaller language editions remain largely understudied and present additional challenges in their own right \citep{nigatu2024low,khatri2022social}.

\paragraph{\textsc{WikiGap} content selection.} We selected a random subset of facts to present from other language editions (Sec.~\ref{sec:wikigap:integration:language_fact} for further detail). This selection could be improved for greater localization through applying a geoprovenance classifier \citep{sen2015barriers}, thereby emphasizing those facts that are sourced from a geographical region relevant to the language edition. This is an important step for future work as our respondents expressed the desire for \textit{localized} Volunteered Geographic Information.

% \paragraph{Quiz-based evaluation design.} 
% As described in Section~\ref{sec:quiz-construction}, our quizzes were generated from facts extracted by \textsc{InfoGap} and presented through the \textsc{WikiGap} interface. This design introduces a potential bias: participants were often asked about information that the system had already highlighted, which may overestimate its utility by favoring fact discovery through the sidebar. A stronger design would have included a balanced mix of questions—half of the questions are answerable via the sidebar and half not—in order to better isolate the extension's added value. We also observed that some multiple-choice items showed stylistic artifacts common in LLM-generated quizzes, such as longer or more detailed correct answers, which could unintentionally cue participants. These limitations do not undermine our main findings but caution against treating quiz scores as a direct measure of discoverability. Future work should refine evaluation protocols by incorporating more rigorous distractor generation and balancing fact- versus non-fact-based questions.

\paragraph{Capacity for misinformation contagion.} Language editions vary considerably in their susceptibility to organized disinformation campaigns \citep{kharazian2024governance}. Thus, there is a risk of spreading misinformation from other language editions into English Wikipedia through \textsc{WikiGap}. \textsc{WikiGap} may thus inadvertently make smaller language editions more attractive for actors carrying out disinformation campaigns, as audience size has previously been proposed as a motivating factor for bad-faith contributions \citep{kharazian2024governance}. Care should be thus taken in deploying \textsc{WikiGap}, with attention paid to the language editions and topics that are selected for the extension.

\paragraph{System Architecture.}
\textsc{WikiGap} demonstrates a socio-technical intervention that aims to reshape how people access, relate to, and interpret multilingual knowledge, which is beyond a purely technical innovation. We have an overwhelmingly positive result from the evaluation. It's important to acknowledge that these results come from a relatively controlled setting. While the findings are encouraging, future work will need to address the technical and interaction challenges of deploying a live, scalable version of the system in more variable real-world conditions.

\paragraph{Future Work.}
Building on our discussion of pluralistic LLM interfaces, future work could extend \textsc{WikiGap}'s design principles to other multilingual information-sharing platforms. Prior work shows that platforms such as Twitter and Flickr also exhibit substantial cross-lingual divergence in how events, entities, and cultural topics are discussed \citep{Graham02102014, mendelsohn2023bridging,10.1504/IJWBC.2010.033753,10.1145/1753326.1753472}. Applying \textsc{WikiGap} could help users interpret how narratives differ across languages and communities beyond Wikipedia. Extending this approach across platforms would also inform the design of more culturally aware LLM interfaces, reinforcing our broader argument that pluralistic system design can mitigate the homogenizing effects of consolidated AI infrastructures.

%% file: sections/7-conclusion.tex
\textsc{WikiGap} contends with longstanding structural inequalities in Wikipedia by enabling readers to access cross-lingual facts directly within the English interface. Through a novel combination of fact-centric text comparison \citep[\textsc{InfoGap};][]{samir-etal-2024-locating} based on LLMs \cite{min2023factscore} and user-centered design, the system unobtrusively displays knowledge missing due to linguistic and editorial asymmetries. Our evaluation shows that \textsc{WikiGap} not only improves usability and fact-finding efficiency, but also fosters greater awareness of cultural variation and the limitations of assuming English Wikipedia as a default knowledge base. By surfacing multilingual content in context, \textsc{WikiGap} connects readers with perspectives on topics that are situated outside of the Anglosphere, connections that were previously occluded by the hegemonic dominance of English Wikipedia \citep[][Chapter 3]{kumar2021digital}.  This work demonstrates how augmentative tools can shift reader practices and challenge epistemic assumptions, paving the way for more inclusive, transparent, and culturally grounded knowledge infrastructures. This paper examines how interface design can challenge the English-as-superset assumption by making multilingual knowledge differences visible to readers during everyday Wikipedia use.

%% file: sections/appendix.tex
\appendix

\section{Sample Quiz Questions}
\label{sec:sample_quiz_questions}
We present two example questions from each of the five topics used in our study. These examples illustrate the format and cultural specificity of the quiz items. All questions were drawn from the 30 language-exclusive facts surfaced by \textsc{WikiGap}, and were manually reviewed for clarity and accuracy. The correct answers are bolded in Table~\ref{tab:quiz-landscape}.

\begin{table}[H]
\centering
\scriptsize
\setlength{\tabcolsep}{3pt}
\caption{Example Quiz Questions Across Topics. Correct answers are bolded.}
\label{tab:quiz-landscape}
% \begin{adjustbox}{width=\linewidth}
% \begin{tabular}{p{3.2cm}p{7.5cm}p{7.5cm}}
\begin{tabular}{p{.12\textwidth}p{.375\textwidth}p{.375\textwidth}}
\toprule
\textbf{Topic} & \textbf{Question 1} & \textbf{Question 2} \\
\midrule

\textbf{Peking Duck} &
\textit{What unique preparation step is done to the duck before roasting it for Peking duck?}  
\newline A. It is marinated in soy sauce  
\newline \textbf{B. It is inflated with air under the skin}  
\newline C. It is stuffed with rice  
\newline D. It is boiled in water 
&
\textit{What is served after the meat in a traditional Peking duck meal?}  
\newline A. A fruit platter  
\newline B. Rice  
\newline \textbf{C. Chinese cabbage soup}  
\newline D. A cup of tea \\ \midrule

\textbf{Wiener Schnitzel} &
\textit{During which historical period was Wiener schnitzel brought to Italy and then to Austria?}  
\newline A. World War I  
\newline B. French Revolution  
\newline \textbf{C. Napoleonic Wars}  
\newline D. Renaissance 
&
\textit{Which city of China has a type of Western cuisine similar to Vienna schnitzel?}  
\newline A. Beijing  
\newline B. Nanjing  
\newline C. Guangdong  
\newline \textbf{D. Shanghai} \\ \midrule

\textbf{Paella} &
\textit{What festival is Spanish paella associated with?}  
\newline A. La Tomatina  
\newline B. San Fermín  
\newline \textbf{C. Falles}  
\newline D. Semana Santa 
&
\textit{How was paella traditionally eaten?}  
\newline A. On plates  
\newline B. In bowls  
\newline \textbf{C. Straight from the cooking pan}  
\newline D. Consumed with bread \\ \midrule

\textbf{Philippine Adobo} &
\textit{Which ingredient is sometimes added to adobo to replace palm or coconut sugar and help tenderize the meat?}  
\newline A. Honey  
\newline B. Molasses  
\newline \textbf{C. Sprite}  
\newline D. Maple syrup 
&
\textit{Which of the following is a traditional method used in adobo to keep meat fresh in tropical climates?}  
\newline A. Drying the meat  
\newline \textbf{B. Frying with vinegar}  
\newline C. Smoking the meat  
\newline D. Freezing the meat \\ \midrule

\textbf{Injera} &
\textit{How large is traditional injera typically?}  
\newline A. 30 centimeters  
\newline B. Half a meter  
\newline \textbf{C. About 1 meter}  
\newline D. 2 meters 
&
\textit{Injera closely resembles which Middle Eastern pancake variant?}  
\newline A. Naan  
\newline \textbf{B. Lahoh}  
\newline C. Pita  
\newline D. Lavash \\

\bottomrule
\end{tabular}
% \end{adjustbox}
\end{table}

\section{Interview and Questionnaire Protocol}
\label{sec:protocol}

Participants were asked the following questions during either the semi-structured interview or the open-ended questionnaire.

\begin{itemize}
    \item Describe your experience using \textsc{WikiGap}. What stood out to you the most, and why?
    \item When you enabled \textsc{WikiGap} to see multilingual facts, how did you feel about the way information was presented? What aspects of the presentation worked well or could be improved?
    \item Can you describe a specific fact you discovered through \textsc{WikiGap} that surprised you or changed your understanding of the topic? How did this discovery influence your perception or understanding of diverse cultural viewpoints?
    \item When using \textsc{WikiGap}, how did you decide which information to trust? Please explain your reasoning.
    \item After using \textsc{WikiGap}, do you see Wikipedia as more reliable or complete, especially for finding information from other languages? Please explain your answer with examples if possible.
    \item In what ways, if any, did \textsc{WikiGap} change your perception or understanding of diverse cultural viewpoints related to the topics you explored?
    \item Did using \textsc{WikiGap} make you more inclined to explore content outside your primary language in the future? Why or why not?
\end{itemize}

\section{Average Completion Time and Accuracy across Different Topics}
\begin{figure}[H]
    \centering
    \includegraphics[width=0.8\linewidth]{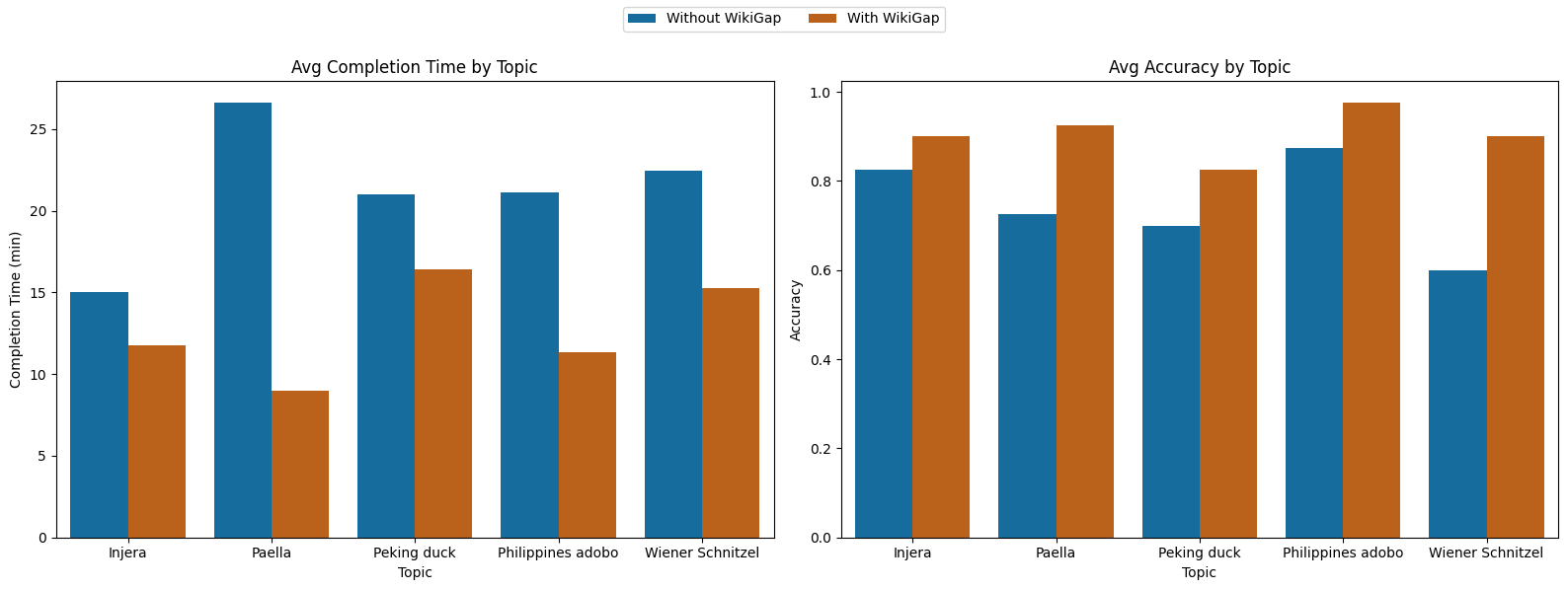}
    \caption{
        Side-by-side bar charts comparing \textbf{average completion time} and \textbf{accuracy} across different topics, with and without the \textsc{WikiGap} extension. \textcolor{darkcerulean}{Blue} represents the control condition (no \textsc{WikiGap}), and \textcolor{burntorange}{orange} represent the treatment condition (with \textsc{WikiGap}). 
    }
    \label{fig:quiz_detailed}
\end{figure}